\documentclass[a4paper,fleqn,review]{cas-sc}

\usepackage[authoryear]{natbib}
\usepackage{siunitx}
\usepackage{subcaption}
\usepackage{placeins}
\usepackage{multirow}
\usepackage{lastpage}
\usepackage{makecell}

\def\tsc#1{\csdef{#1}{\textsc{\lowercase{#1}}\xspace}}
\tsc{WGM}
\tsc{QE}
\tsc{EP}
\tsc{PMS}
\tsc{BEC}
\tsc{DE}

\makeatletter
\g@addto@macro\ps@first{%
  \def\@oddfoot{\hbox to \textwidth{\hfil\footnotesize\itshape Page \thepage\ of \pageref{LastPage}}}%
  \let\@evenfoot\@oddfoot
}
\g@addto@macro\ps@cas{%
  \def\@oddfoot{\hbox to \textwidth{\hfil\footnotesize\itshape Page \thepage\ of \pageref{LastPage}}}%
  \let\@evenfoot\@oddfoot
}
\makeatother

\begin{document}
\let\WriteBookmarks\relax
\let\printorcid\relax
\def\floatpagepagefraction{1}
\def\textpagefraction{.001}
\shorttitle{Avalanche Interaction with Obstacle Arrays}
\shortauthors{Sharma and Bhutani}

\title [mode = title]{Numerical Investigation of Avalanche Interaction with Obstacle Arrays: Effects of Obstacle Height, Width, and Shape on Flow Dynamics}                      

\author[1]{Vishal Sharma}
\ead{d22051@students.iitmandi.ac.in}

\credit{Writing--original draft, Visualizations, Validation, Software, Investigation, Methodology, Conceptualization}

\affiliation[1]{organization={School of Mechanical and Materials Engineering, Indian Institute of Technology Mandi},
                city={Mandi},
                postcode={175075}, 
                state={Himachal Pradesh},
                country={India}}

\author[1]{Gaurav Bhutani}[%
   ]
\cormark[1]
\ead{gaurav@iitmandi.ac.in}

\credit{Writing--review \& editing, Supervision, Resources, Formal analysis, Conceptualization}

\cortext[cor1]{Corresponding author}

\begin{abstract}
Snow avalanches pose a significant threat to infrastructure in mountainous regions, motivating the development of effective flow mitigation measures. This study numerically investigates the interaction of granular avalanches with obstacle arrays, focusing on the effects of obstacle height, width, and shape on flow dynamics and obstruction efficiency. A depth-averaged finite-volume model employing Mohr--Coulomb rheology is validated against laboratory experiments and subsequently used to examine arrays with uniform and variable obstacle dimensions, as well as uniform and mixed-shape configurations. The simulations capture bow-shock formation, upstream deposition, flow diversion, and downstream grain-free regions. Under a constant obstacle-volume constraint, increasing obstacle width relative to height enhances flow obstruction by reducing the effective gaps between obstacles, with uniform obstacle dimensions providing better flow restriction than the variable-height configurations examined. Comparisons of cylindrical, thin cuboid, and diamond cuboid obstacles reveal that diamond-shaped obstacles in the leading row produce greater upstream deposition and blockage while restricting downstream flow propagation. However, these configurations also experience higher impact forces, highlighting a trade-off between mitigation effectiveness and structural loading. Empirical correlations are developed to predict the maximum flow depth, its location, and the corresponding mean flow velocity based on obstacle geometry. Application to the CT-8 avalanche site in the Indian Himalayas demonstrates approximately 20\% and 25\% reductions in roadside flow depth and mean velocity, respectively. The findings provide insights into avalanche--structure interactions and support the geometric design of obstacle arrays for avalanche mitigation.
\end{abstract}

\begin{keywords}
Snow avalanche mitigation \sep
Obstacle arrays \sep
Granular flow dynamics \sep
Depth-averaged modelling \sep
Flow obstruction
\end{keywords}

\maketitle

\section{Introduction}\label{sec1}

Avalanches are one of the many natural hazards in mountainous regions, particularly in areas with human settlements and infrastructure. In regions like the Western Himalayas, frequent avalanches pose a persistent threat to lives and property. Effective mitigation strategies are essential to reduce their impact. Avalanche mitigation typically involves the use of defense measures designed to divert, slow down, or arrest the flow. These measures are primarily classified as active and passive defense measures. Active measures aim to control avalanches from starting or directly influence the flow. Among these, arrays of obstacles have gained popularity due to their lower cost, ease of installation, and effectiveness in energy dissipation. When strategically placed along the flow path of an avalanche, these arrays can disrupt avalanche momentum, alter flow patterns, and reduce runout distances, especially when arranged in multiple rows. 

Previous studies have demonstrated that obstacle geometry, spacing, and arrangement strongly influence avalanche and granular flow behavior \citep{faug2002experimental,hauksson2007laboratory,zanuttigh2006experimental,sheikh2008interaction}. More recent investigations have shown that staggered and multiple-row obstacle arrays provide greater energy dissipation and flow control than single obstacles, highlighting their potential for avalanche mitigation \citep{hakonardottir2003laboratory,naaim2003dry,jianbo2020experimental,luong2020spread,darbois2023downslope}. Depth-averaged (DA) models have become a widely adopted framework for simulating avalanche and granular flow behavior, offering a balance between computational efficiency and physical accuracy. These models have been successfully applied to investigate obstacle-induced flow diversion and shock-wave formation \citep{cui2014computational,fei2020modelling,tregaskis2022subcritical}. More recently, DA models have been used to evaluate obstacle array performance, including staggered pile configurations, forested slopes represented as dense cylindrical arrays, and the influence of obstacle density and geometry on flow deflection, velocity reduction, and deposition \citep{jianbo2020experimental,luong2020spread,darbois2023downslope,wang2022runout}.
Despite these advances, most previous studies have focused on obstacle arrays with uniform obstacle height and shape at a laboratory scale. Consequently, the influence of variable obstacle heights, as well as mixed obstacle shapes within an array, remains poorly understood.

In the present study, an open-source depth-averaged solver is employed to investigate the influence of obstacle array configuration on avalanche dynamics. The numerical model is first validated against laboratory experiments \citep{caviedes20142d} and subsequently used to examine arrays with uniform and variable obstacle heights, and uniform and mixed obstacle shape configuration. The proposed configurations are then applied on the CT-8 avalanche site in the Indian Himalayas using a high-resolution digital elevation model to evaluate their effectiveness under realistic terrain conditions. The results provide new insights into the design of obstacle arrays for avalanche mitigation and demonstrate the applicability of depth-averaged modeling for assessing their performance over natural terrain.

The remainder of this paper is organized as follows: Section~\ref{sec3} outlines the governing equations and numerical methodology. Section~\ref{sec4} presents the validation results against published data. Section~\ref{sec5} discusses the simulation results for arrays with uniform and variable obstacle heights and widths, followed by the discussion on arrays with uniform and mixed-shape obstacle configurations. This is followed by the inclusion of the empirical correlations obtained from the numerical data to relate flow parameters with the obstacle height, width and shape. Finally, the application of obstacle arrays to a real avalanche site is presented, and Section~\ref{sec:7} concludes the paper with key findings and limitations.

\section{Methodology}\label{sec3}
This section describes the numerical approach used in the present study to model granular avalanches and their interaction with obstacles. It begins with the formulation of the depth-averaged governing equations and the underlying stress assumptions adopted in this framework. This is followed by a description of the Mohr–Coulomb rheological model used in this work and the corresponding source terms. Finally, the section outlines the prescribed initial and boundary conditions, along with the finite-volume method implemented in \texttt{TITAN2D} with adaptive mesh refinement.

\subsection{Depth-averaged equations}\label{sec3sub1}
Avalanche flow is modeled using a depth-averaged incompressible continuum approach. In this approach, averaging over the flow depth reduces the three-dimensional governing equations to a two-dimensional form, making them more tractable for numerical solution. The depth-averaged continuity equation is given as:
\begin{equation} \frac{\partial h}{\partial t} + \frac{\partial }{\partial x}(h \bar{u}) +\frac{\partial }{\partial y}(h \bar{v}) =0. \label{eq:continuity} \end{equation}
Here, $h(x,y,t)$ represents the flow depth measured along the $z$-direction, which is normal to the surface in a curvilinear (local) orthogonal coordinate system. The coordinate $x$ denotes the downslope direction, while $y$ is perpendicular to $x$ in the flow plane.
In this paper, the terms \textit{flow depth} and \textit{flow height} are used interchangeably to refer to \( h \).
The variables $\bar{u}$ and $\bar{v}$ are the depth-averaged velocity components in the $x$- and $y$-directions, respectively. The overbar denotes depth-averaged variables, defined as the mean value of a quantity over the flow depth.

In the depth-averaged model, the vertical normal stress, \(\sigma_{zz}\), is assumed to be hydrostatic, and expressed as:
\begin{equation}
	\sigma_{zz}(z) = \rho g_z (h - z),
	\label{eq:normal_stress}
\end{equation}
where \(g_z\) is the component of gravity vector along $z$-direction. The normal stresses in the downslope (\(x\))- and cross-slope (\(y\))-directions are related to the vertical normal stress by the earth pressure coefficient, \(k\), such that:
\begin{equation}
	\bar{\sigma}_{xx} = k\,\bar{\sigma}_{zz}, \qquad \bar{\sigma}_{yy} = k\,\bar{\sigma}_{zz}.
	\label{eq:lateral_stresses}
\end{equation}

For Coulomb-type closure, the earth pressure coefficient is expressed in terms of the internal friction angle $\phi$ and bed friction angle $\delta$ as:
\begin{equation}
	k = 2 \left[ \frac{1 \pm [|1-\cos^2{\phi}(1 + \tan^2{\delta})|]^{1/2}}{\cos^2{\phi}} \right] - 1.
	\label{eq:scalar_k}
\end{equation} 

The sign ($\pm$) depends on whether the material is stretching or being compressed, corresponding to active and passive stress states, respectively. The negative sign represents the active state, defined by \( \frac{\partial \bar{u}}{\partial x} + \frac{\partial \bar{v}}{\partial y} > 0 \), while the positive sign corresponds to the passive state, defined by \( \frac{\partial \bar{u}}{\partial x} + \frac{\partial \bar{v}}{\partial y} < 0 \). Furthermore, to maintain the numerical stability of the solver and ensure real, positive values of $k$, the absolute value of the expression under the square root was employed.

Basal resistance is represented using a dry Coulomb friction law, under the assumption that the flow continuously yields at the base. Accordingly, the basal shear stress term along the downslope direction is given by \(\sigma_{xz}^b = \sigma_{zz}^b \tan\delta\), where $\sigma_{zz}^b$ represents basal normal stress. In addition, the effect of basal surface curvature is accounted for through a centripetal acceleration term. As a result, the total basal normal stress in downslope direction includes contributions from both the hydrostatic and the centripetal acceleration, and is expressed as:
\begin{equation}
	\sigma_{total}^b
	=
	\rho h g_z
	+
	\rho h\,\frac{\bar{u}^2}{r_x}.
	\label{eq:total_stress}
\end{equation}

The internal shear stress $\sigma_{xy}$ formulation is obtained using a Coulomb yield criterion and Mohr-circle representation, written as:
\begin{equation}
	\bar{\sigma}_{xy}
	=
	\mathrm{sgn}\!\left(\frac{\partial \bar{u}}{\partial y}\right)\,
	\bar{\sigma}_{xx}\,\sin\phi.
	\label{eq:internal_stress}
\end{equation}

Substituting the expressions for basal and internal stresses into the depth-averaged momentum equations yields the governing equations in the \(x\)- and \(y\)-directions, consistent with the formulation of \citet{patra2005parallel}. In this formulation, the bulk density appears as a common multiplicative factor in all terms on both sides of the equations. As a result, the density $\rho$ cancels out in the final conservative form, and the computed flow variables become independent of $\rho$. The resulting two-dimensional depth-averaged governing equations are given as:
\begin{equation}
	\frac{\partial }{\partial t}(h \bar{u}) + \frac{\partial }{\partial x}\left(h \bar{u}^2 + \frac{k g_z h^2}{2}\right) + \frac{\partial }{\partial y}(h \bar{u} \bar{v}) = S_x, 
	\label{eq:x_mom}
\end{equation}
\begin{equation}
	\frac{\partial }{\partial t}(h \bar{v}) + \frac{\partial }{\partial x}(h \bar{u} \bar{v}) + \frac{\partial }{\partial y}\left(h \bar{v}^2 + \frac{k g_z h^2}{2}\right) = S_y, 
	\label{eq:y_mom}
\end{equation}
where \(S_x\) and \(S_y\) represent the source terms, which include the gravitational acceleration, basal friction resisting the flow, and any additional forces arising from specific rheological assumptions. 

For the Mohr--Coulomb (MC) rheology, the source terms $S_x$ and $S_y$ are given by:
\begin{equation}
	S_x = g_x h - \frac{\bar{u}}{|\mathbf{\bar{u}}|} \left[ h \left( g_z + \frac{\bar{u}^2}{r_x} \right) \tan{\delta} \right] 
	- h\, k \, \text{sgn} \left( \frac{\partial \bar{u}}{\partial y} \right) \sin{\phi} \, \frac{\partial (g_z h)}{\partial y},
\end{equation}

\begin{equation}
	S_y = g_y h - \frac{\bar{v}}{|\mathbf{\bar{u}}|} \left[ h \left( g_z + \frac{\bar{v}^2}{r_y} \right) \tan{\delta} \right]
	- h\, k \, \text{sgn} \left( \frac{\partial \bar{v}}{\partial x} \right) \sin{\phi} \, \frac{\partial (g_z h)}{\partial x}.
\end{equation}
Here, \(\mathbf{\bar{u}} = (\bar{u}, \bar{v})\) denotes the depth-averaged velocity vector, and \(r_x\), \(r_y\) are the radii of curvature of the basal surface in the \(x\)- and \(y\)-directions, respectively, expressed in terms of the derivatives of basal surface topography. See \citet{sharma2026computational} for detailed derivation.

\subsection{Initial and boundary conditions}
The initial conditions for the depth-averaged equations are specified by defining the flow depth \(h(x, y)\) and the velocity field \(\mathbf{\bar{u}}(x, y)\) at time \(t = 0\). The flow is initialized using a localized granular pile for the depth, while the initial velocity is typically set to zero. Zero normal mass flux boundary conditions are imposed along all domain boundaries to prevent material from entering or leaving the computational domain during the simulation. Furthermore, the derivation of the depth-averaged equations assumes no mass flux through either the basal surface or the free surface, implying that the terrain is impermeable and that there is no entrainment or detrainment of material.

\subsection{Solution method}\label{sec3sub2}
The governing depth-averaged equations (Equations~\eqref{eq:continuity}, \eqref{eq:x_mom}, and \eqref{eq:y_mom}) were solved using a finite-volume approach with a first-order Godunov scheme. This explicit scheme effectively captures shock-like features, including steep fronts and rapid variations in flow depth. Numerical stability was maintained by enforcing a Courant--Friedrichs--Lewy (CFL) number below 0.5 through appropriate selection of the time-step size. Simulations were performed using the open-source code \texttt{TITAN2D} \citep{patra2005parallel}.

To improve computational efficiency while maintaining accuracy, adaptive mesh refinement (AMR) was employed. This approach dynamically refines the grid in regions with large gradients in flow depth, while using coarser grids in less active regions. As a result, higher resolution is achieved where needed for accurate predictions, while minimizing the overall computational cost. All simulations were performed on a workstation equipped with an 8-core Intel i7-9700K CPU (3.6 GHz) and 64 GB of RAM. The code is parallelized using OpenMP, and the simulations were executed using 8 threads. The combination of a depth-averaged formulation with a parallelized, AMR-based finite-volume method provides a computationally efficient and robust framework for simulating granular avalanche dynamics and their interaction with obstacles over both synthetic and real mountain topographies. It should be noted that AMR was applied only for the real terrain simulations, whereas the small-scale simulations were performed without mesh refinement.

\section{Validation}\label{sec4}
\begin{figure}[htbp]
	\begin{subfigure}[t]{0.48\textwidth}
	\centering
    \includegraphics[width=0.60\textwidth]{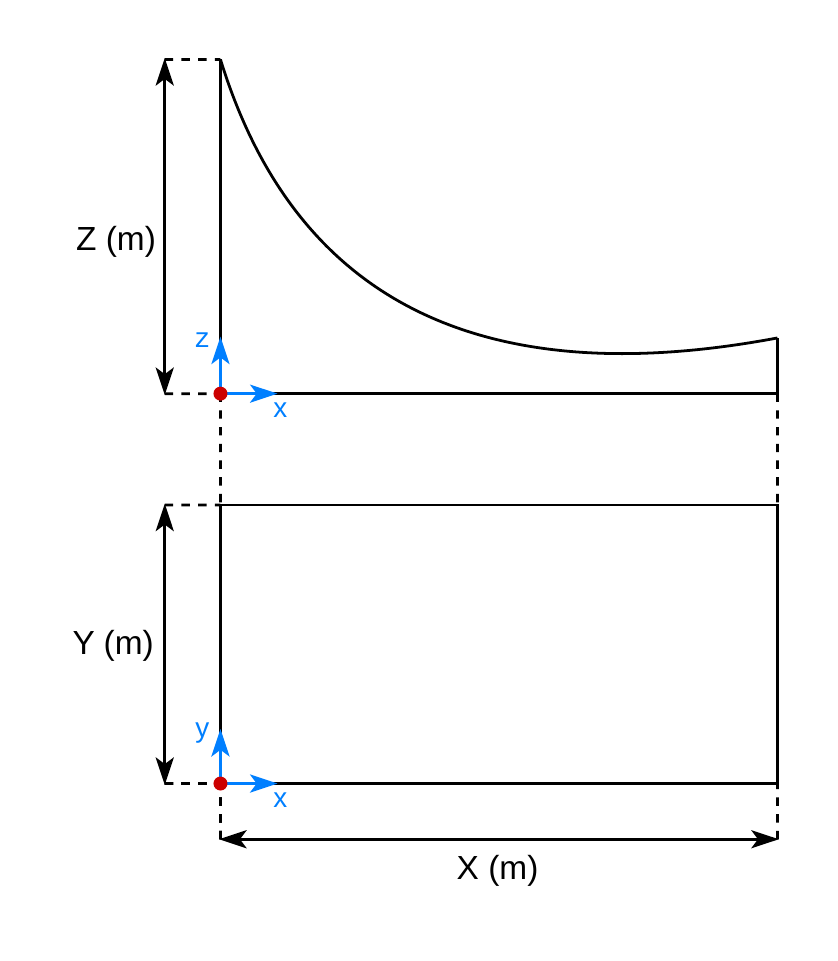}
    \caption{}
    \label{subfig:curved_domain}
	\end{subfigure}
    \hfill
	\begin{subfigure}[t]{0.48\textwidth}
    \centering
	\includegraphics[scale=0.38]{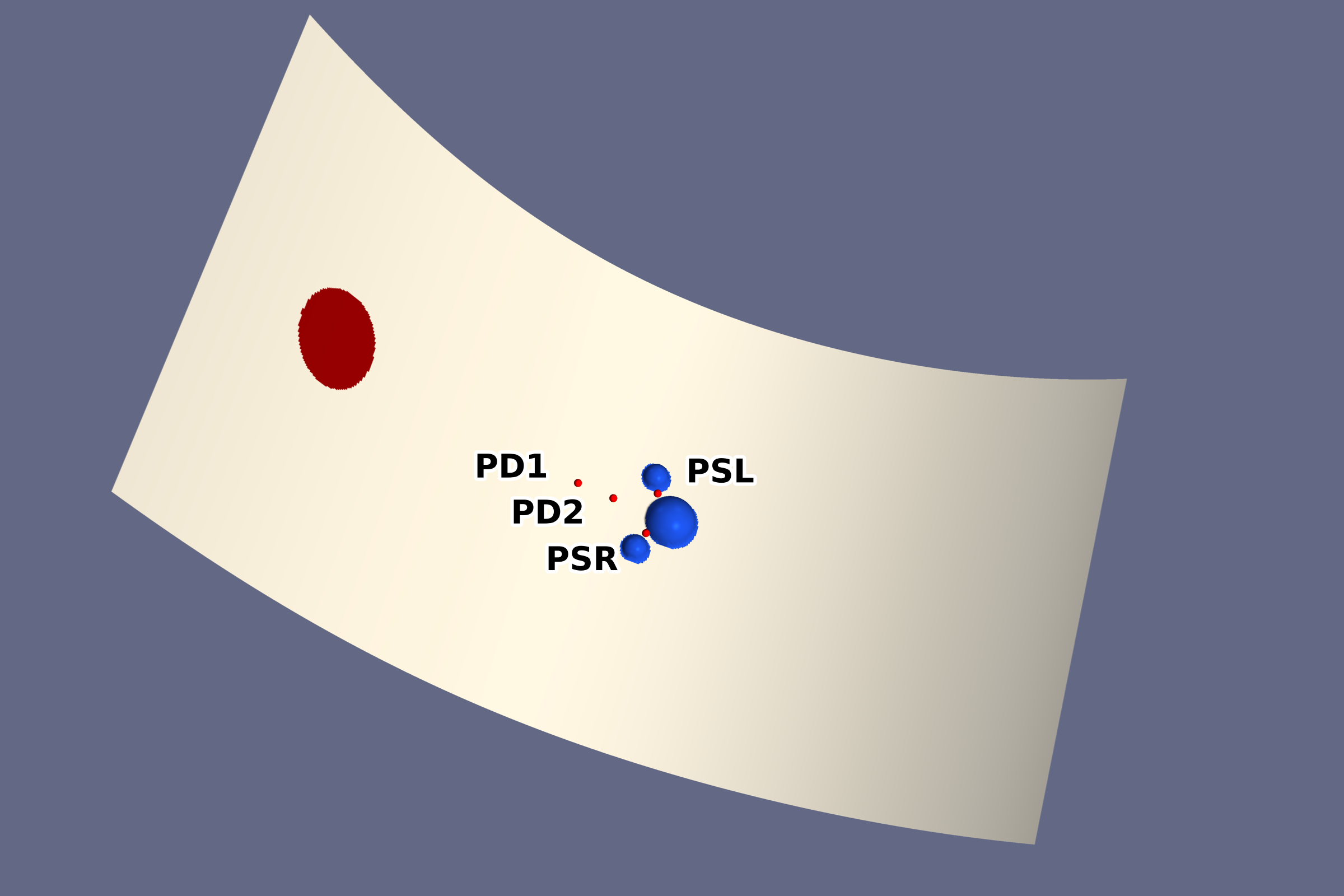}
    \caption{}
    \label{subfig:curved_domain_isometric}
    \end{subfigure}
	\caption{Computational domain of the curved bed profile featuring three hemispherical obstacles. (a) Side and top views of the domain (shown at the top and bottom, respectively), with the origin (red circle) and coordinate directions (blue arrows). (b) Isometric projection of the domain, where the dark red ellipse indicates the projected initial release position of the granular avalanche. The red dots mark the probe locations for measuring flow depth, following the experimental setup of \citet{caviedes20142d}.}
	\label{fig:three_semisphere_obstacle}
\end{figure}

The computational method used in the present study was validated against the experimental work of \citet{caviedes20142d}. The model was used to simulate the interaction of a granular material with multiple hemispherical obstacles configuration. In the reference experimental setup, probes were inserted on a curved surface to measure flow depth at multiple locations. The positions of these probes are listed in Table~\ref{table0} and marked with red dots in Figure~\ref{fig:three_semisphere_obstacle}. 

\begin{table}[width=.9\linewidth,cols=3,pos=h]
	\caption{Probe locations inserted on the curved surface to measure flow depth.}\label{table0}
	\begin{tabular*}{\tblwidth}{@{} LLL@{} }
		\toprule
		Probe & X (\si{\milli\meter}) & Y (\si{\milli\meter}) \\
		\midrule
		PD1 & 600 & 500  \\
		PD2 & 680 & 500  \\
		PSL & 770 & 550  \\
		PSR & 770 & 450  \\
		\bottomrule
	\end{tabular*}
\end{table}

The shape of this curved surface was obtained by fitting a fourth-degree polynomial, as discussed by \citet{caviedes20142d}. This numerical domain of 1.75 \si{\meter} $\times$ 1.0 \si{\meter} was projected on $xy$-plane with a surface structured (square) grid resolution of 0.0028 \si{\meter} as shown in Figure~\ref{subfig:curved_domain}. The granular material used was sand having an effective density ($\rho$) of 1550~\si{\kilogram\per\cubic\meter}, with an internal friction angle taken as 34$^\circ$ based on the study by \citet{juez20142d}. 

The initial condition was consistent across all simulations. The granular avalanche was triggered by the sudden release of a hemispherical cap full of sand. This release pile had a radius of 104 \si{\milli\meter}, which corresponds to a volume that matched with the experiments, and its center located at $x = 100$ \si{\milli\meter}, $y = 500$ \si{\milli\meter}, and $z = 736$ \si{\milli\meter} represented in Figure~\ref{subfig:curved_domain_isometric}. 

\begin{figure}[htbp]
	\centering
	\begin{tabular}{c@{\hspace{0.8cm}}c}
		
		\includegraphics[scale=0.68]{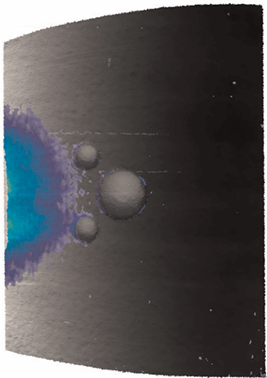} &
		\includegraphics{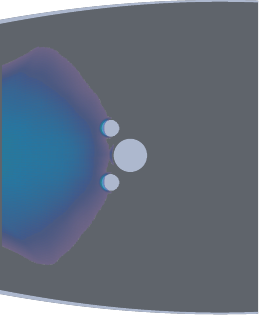} \\[0.5em]
		
		\includegraphics[scale=0.68]{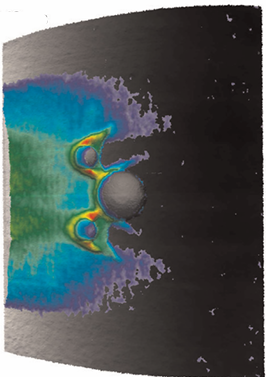} &
		\includegraphics{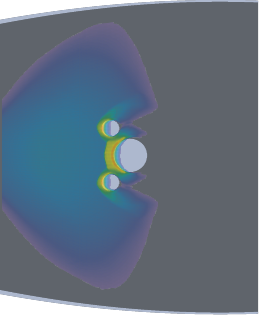} \\[0.5em]
		
		\includegraphics[scale=0.68]{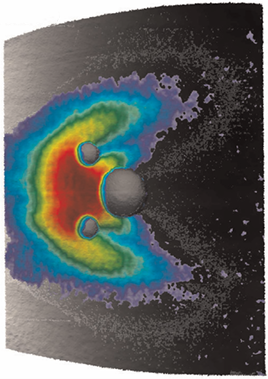} &
		\includegraphics{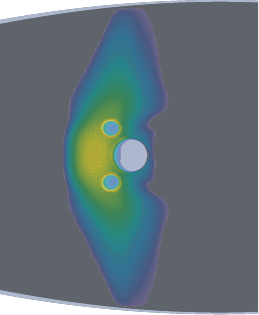}
		
	\end{tabular}
	\vspace{0.2cm}
	
	\includegraphics{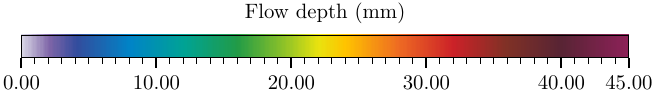}
	
	\caption{Qualitative comparison of the avalanche front and depth contours obtained from experiments (left) and simulations (right) for bed friction angle 33$^\circ$ at time instances $t = 0.4$ \si{\second}, $t = 0.6$ \si{\second}, and $t = 1.5$ \si{\second}.}
	\label{fig:height_contour_bed_33}
\end{figure}

\begin{table}[width=.9\linewidth,cols=5,pos=h]
	\caption{Dimensions and spatial coordinates of the three hemispherical obstacles.}\label{table1}
	\begin{tabular*}{\tblwidth}{@{} LLLLL@{} }
		\toprule
		Obstacle & Diameter (\si{\milli\meter}) &  X (\si{\milli\meter}) & Y (\si{\milli\meter}) & Z (\si{\milli\meter}) \\
		\midrule
		Large & 59 & 814 & 500 & 207  \\
		Small-1 & 29 & 757 & 410 & 233 \\  
        Small-2 & 29 & 757 & 590 & 233 \\
		\bottomrule
	\end{tabular*}
\end{table}
Downstream of the release, a configuration comprising three hemispherical obstacles was evaluated, with its dimensions and spatial coordinates summarized in Table~\ref{table1}. Before comparing the numerical results with the experimental data, a sensitivity analysis of the bed friction angle ($\delta$) was conducted. This step was necessary because this parameter could not be directly measured and was not provided in the referenced experimental study. Consequently, the analysis was performed by varying $\delta$ from 26$^\circ$ to 36$^\circ$. It was observed that at $\delta = 26^\circ$, the granular mass completely overflowed the obstacle because of the lower basal resistance experienced by the flow. Conversely, at $\delta = 36^\circ$, the flow arrested immediately after interacting with the obstacle. To accurately capture the granular movement, a qualitative comparison of the avalanche front was conducted between the numerical and experimental data. As demonstrated in Figure~\ref{fig:height_contour_bed_33}, a bed friction angle of $\delta = 33^\circ$ provided the most accurate representation of the experimental avalanche front, and was therefore adopted for all subsequent simulations in this Section.

\FloatBarrier

\begin{figure}[htbp]
	\begin{subfigure}[t]{0.48\textwidth}
		\centering
		\includegraphics[width=\textwidth]{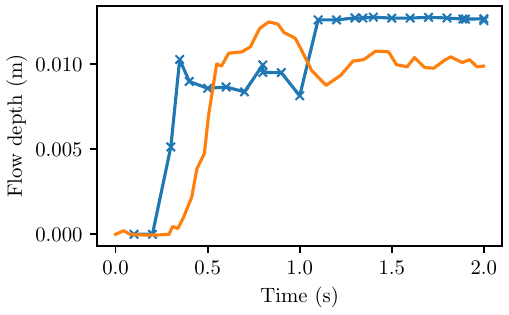}
		\caption{PD1} 
		\label{fig:subfig_three:pd1}
	\end{subfigure}
	\hfill 
	\begin{subfigure}[t]{0.48\textwidth}
		\centering
		\includegraphics[width=\textwidth]{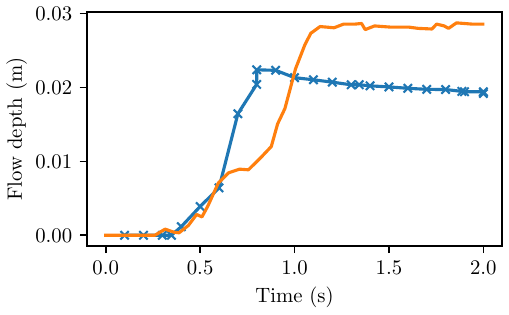}
		\caption{PD2}
		\label{fig:subfig_three:pd2}
	\end{subfigure}
	
	\vspace{1em}
	
	\begin{subfigure}[t]{0.48\textwidth}
		\centering
		\includegraphics[width=\textwidth]{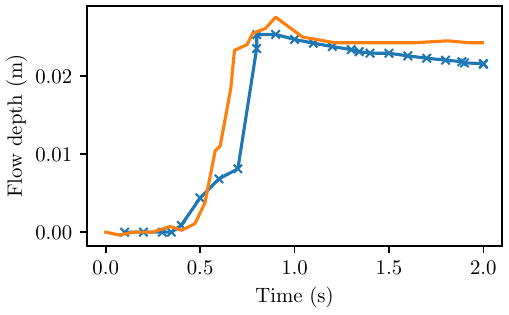}
		\caption{PSL}
		\label{fig:subfig_three:psl}
	\end{subfigure}
	\hfill
	\begin{subfigure}[t]{0.48\textwidth}
		\centering
		\includegraphics[width=\textwidth]{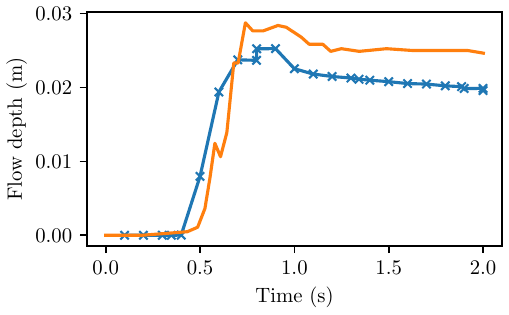}
		\caption{PSR}
		\label{fig:subfig_three:psr}
	\end{subfigure}
	\caption{Comparison of computed results with experimental data \citep{caviedes20142d}. Solid line: experimental results; solid line with crosses: present work.}
	\label{fig:compare_three_semisphere_data}
\end{figure}

The numerical data obtained at all probe locations were compared against the experimental measurements, as illustrated in Figure~\ref{fig:compare_three_semisphere_data}. Probes PD1 and PD2 are located upstream of the hemispherical obstacles configuration to capture upstream shock-wave formation and flow deposition, whereas probes PSL and PSR are positioned laterally to the central hemispherical obstacle to observe the evolution of the flow depth in its immediate vicinity. 
At probe PD1, the numerical simulation predicted an earlier arrival of the granular front and an earlier onset of the flow-depth jump than observed experimentally. Nevertheless, the initial rate of increase in flow depth and the subsequent quasi-steady depth were in good agreement with the measurements.
Furthermore, the flow depth observed during the steady state closely matched the experimental values. Similarly, across the remaining probe locations, the time of jump, the temporal slope of the flow depth, and the steady-state depth values were all closely aligned with the experimental measurements. Ultimately, the present model successfully captured the temporal evolution of the flow depth at multiple probe locations, demonstrating strong agreement with the experimental data. 

Following this validation, the numerical method is applied to analyze interactions between full obstacle arrays and granular avalanche flows. Specifically, the next section details how flow dynamics are influenced by variable obstacle heights and mixed geometric configurations.

\section{Results and discussion}\label{sec5}
This section presents the simulation results for avalanche interaction with obstacle arrays featuring varying heights and mixed geometric shapes. First, the qualitative analysis of granular avalanche interaction with obstacle array is discussed, followed by the identification of the effective array configurations based on key performance metrics, such as the blockage ratio and velocity reduction ratio, which are evaluated and compared. Subsequently, empirical correlations based on simulation data across all array configurations are introduced to predict the maximum flow depth, its spatial location, and the corresponding mean flow velocity. Finally, the array configurations are applied to a real-world mountain avalanche site to evaluate their impact on avalanche flow parameters.

\begin{figure}[htbp]
	\centering
	\includegraphics[scale=0.5]{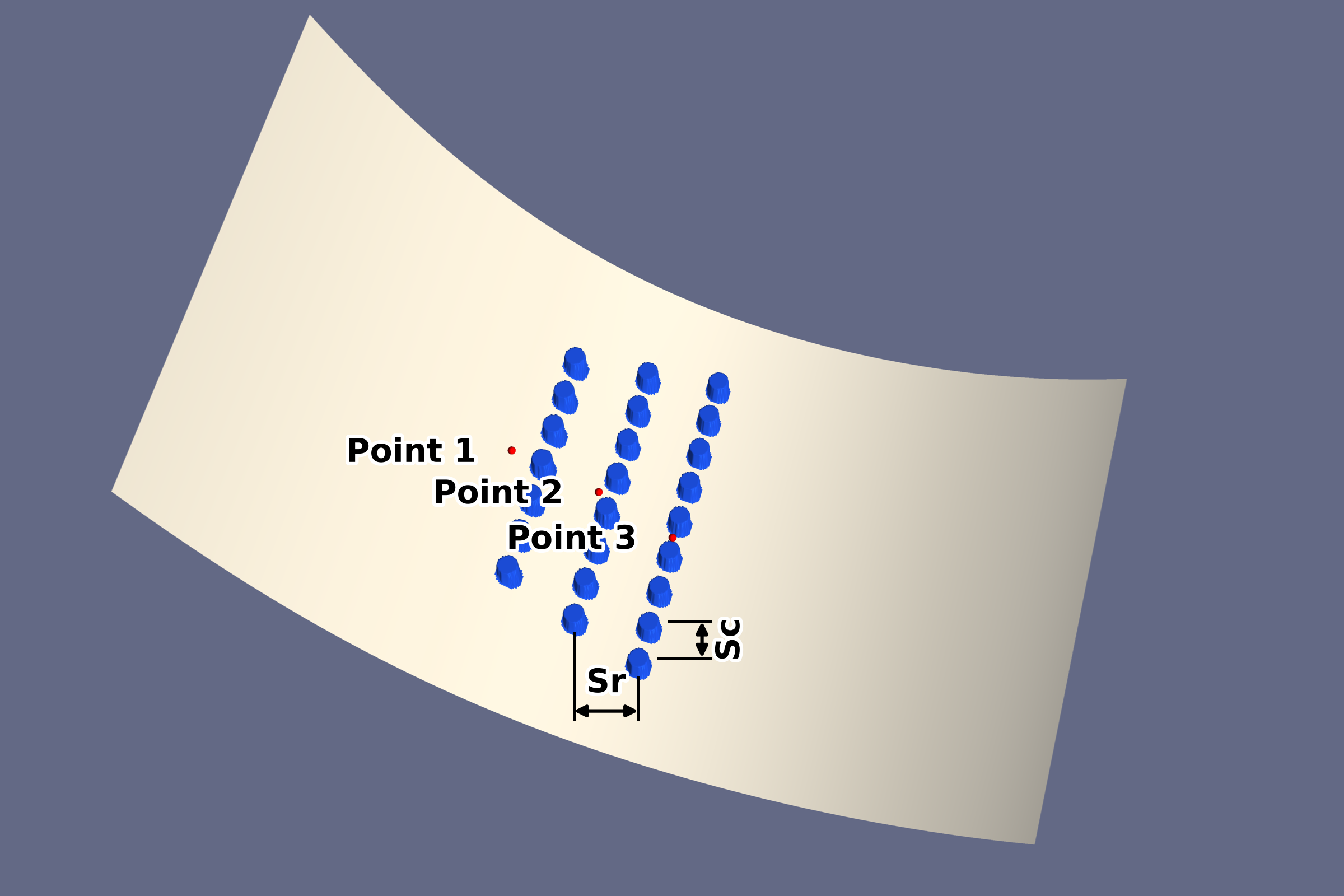}
	\caption{Array of cylindrical obstacles (blue color) positioned on a synthesized curved surface. Red dots indicate the three measurement locations. The labels Sr and Sc denote the spacing between consecutive rows and the distance between adjacent obstacles within a row, respectively.}
	\label{fig:array_cylinder}
\end{figure}

The optimal three-row obstacle configuration layout utilized in this study---comprising 7, 8, and 9 obstacles in its consecutive rows, respectively---is inspired by the work of \citet{bi2022optimal}. Furthermore, the obstacle configuration with inter-obstacle spacing ($S_c$) of 0.0875 \si{\meter} and inter-row spacing ($S_r$) of 0.1575 \si{\meter} was chosen based on the arguments of optimal staggered layout \citep{bi2022optimal} as shown in Figure~\ref{fig:array_cylinder}. 

\subsection{Flow analysis of avalanche-array impact}
\begin{figure}[htbp]
	\centering
	\begin{tabular}{c@{\hspace{1.0cm}}c}
		
		\includegraphics{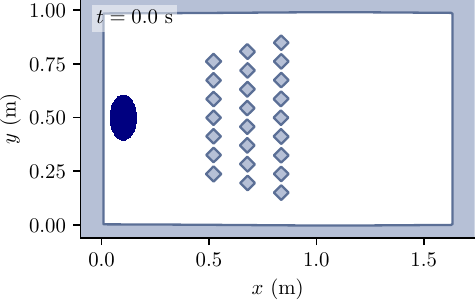} &
		\includegraphics{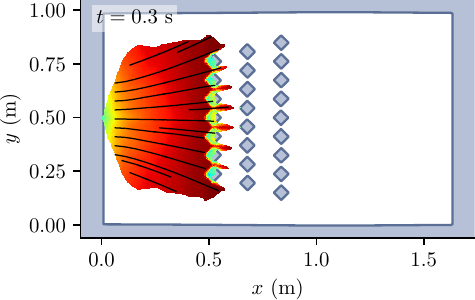} \\[0.5em]
		
		\includegraphics{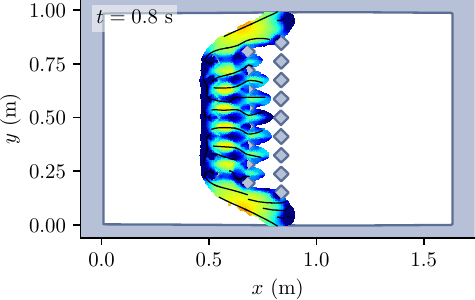} &
		\includegraphics{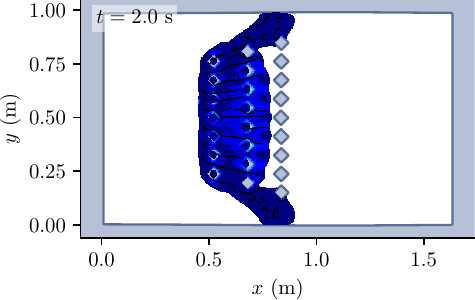} \\[0.5em]
		
	\end{tabular}
	\vspace{0.2cm}
	
	\includegraphics{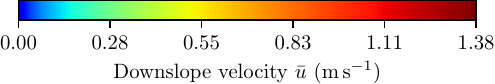}
	
	\caption{Depth-averaged downslope velocity $\bar{u}$ of the avalanche impacting array of diamond cuboid obstacles with obstacle height 0.06 \si{\meter} and width 0.048 \si{\meter} at time instances $t = 0.0$ \si{\second}, $t = 0.3$ \si{\second}, $t=0.8$ \si{\second}, and $t = 2.0$ \si{\second}. Streamlines, calculated from instantaneous velocity field are overlaid with black lines. The diamond cuboid obstacles are highlighted in gray, and grain-free region is shown in white.}
	\label{fig:downslope_velocity_array_cylinder}
\end{figure}

\begin{figure}[htbp]
	\centering
	\begin{tabular}{c@{\hspace{0.1cm}}c}
		
		\includegraphics{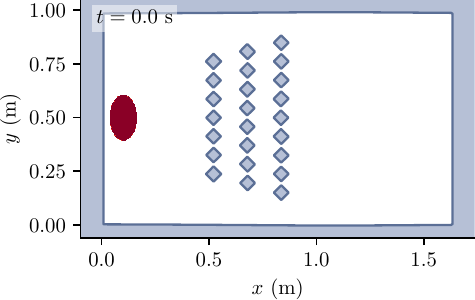} &
		\includegraphics{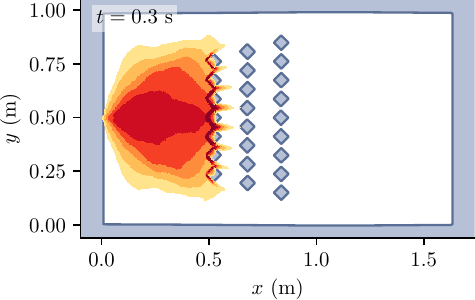} \\[0.5em]
		
		\includegraphics{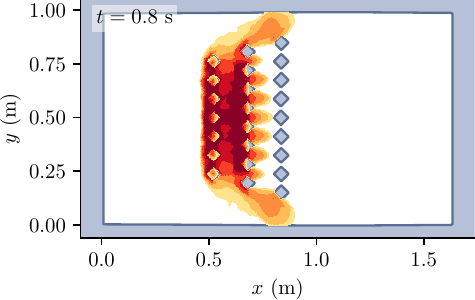} &
		\includegraphics{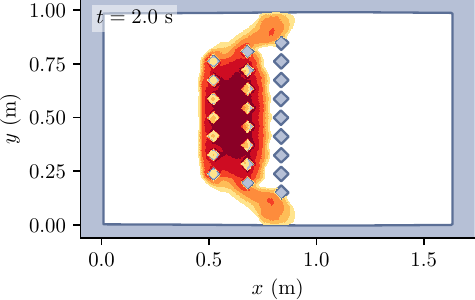} \\[0.5em]
		
	\end{tabular}
	\vspace{0.2cm}
	
	\includegraphics{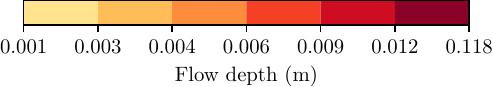}
	
	\caption{Flow depth ($h$) contours of the avalanche impacting array of diamond cuboid obstacles with obstacle height 0.06 \si{\meter} and width 0.048 \si{\meter} at time instances $t = 0.0$ \si{\second}, $t = 0.3$ \si{\second}, $t=0.8$ \si{\second}, and $t = 2.0$ \si{\second}. The diamond cuboid obstacles are highlighted in gray, and grain-free region is shown in white.}
	\label{fig:height_contours_array_cylinder}
\end{figure}

When a granular avalanche interacts with an obstacle, the material diverts around it, significantly altering both the localized flow depth and velocity. This interaction results in the formation of a bow shock upstream of a static deposit (dead zone), while a distinct grain-free region develops downstream \citep{Gray2003shockwaves}. As illustrated in Figures~\ref{fig:downslope_velocity_array_cylinder} and \ref{fig:height_contours_array_cylinder}, the spatial contours for the uniform diamond cuboid configuration (later referred to as the ADC-OH1 case in Table~\ref{tab:numerical_data}) clearly reveal the distinct features of this interaction. Upon initial impact of the avalanche with the first row ($t \approx 0.3\,\si{\second}$), the rapidly flowing mass abruptly decelerated to form a prominent bow shock, driving material accumulation upstream of the obstacles. This accumulated material came to rest, creating a dead zone that propagated upstream over time. As the remaining incoming flow interacted with this dead zone, the previously deposited material was forced to divert around the obstacles at a minimal velocity ($t \approx 0.8\,\si{\second}$). Interacting with the second row, this diverted flow faced further obstruction, resulting in substantial deposition between the first and second rows ($t = 2.0\,\si{\second}$). Ultimately, continuous energy dissipation across the entire array severely restricted the material reaching the third row.

\begin{figure}[htbp]
	\centering
	\includegraphics{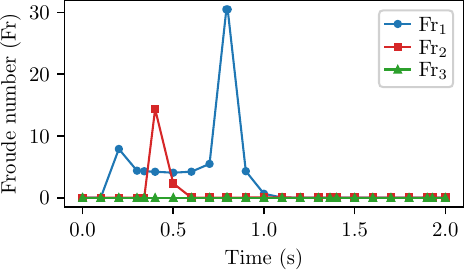}
	\caption{Variation of Froude number with time for a granular avalanche interacting with an array of diamond cuboid obstacles with obstacle height 0.06 \si{\meter} and width 0.048 \si{\meter}. $Fr_1$, $Fr_2$, and $Fr_3$ shows the Froude number corresponding to Point 1 (upstream of an array), Point 2 (intermediate), and Point 3 (downstream of an array) shown in Figure~\ref{fig:array_cylinder}.}
	\label{fig:froude_number}
\end{figure}

To quantitatively characterize these observed flow regime transitions, the Froude number ($Fr$) was evaluated at three critical locations across the array (Figure~\ref{fig:array_cylinder}): upstream of the first row, upstream of the second row, and near the third row. The Froude number dictates the localized flow state---supercritical ($Fr > 1$), critical ($Fr = 1$), or subcritical ($Fr < 1$)---and is defined as:
\begin{equation}
Fr = \frac{|\mathbf{\bar{u}}|}{\sqrt{gh\cos\zeta}},
\end{equation}
where $\zeta$ represents the slope inclination angle. As depicted in Figure~\ref{fig:froude_number}, the temporal evolution of $Fr$ mirrors the physical phenomena observed in the spatial contours. Initially, before the granular material impacted the first row ($t \approx 0.3\,\si{\second}$), the upstream flow remained highly supercritical ($Fr_1 > 1$). However, after the impact with the first row a sudden formation of the bow shock triggered a drastic drop in $Fr_1$. As this shock propagated upstream to form the dead zone, kinetic energy rapidly dissipated, keeping $Fr_1$ minimal until $t \approx 0.8\,\si{\second}$. As the remaining incoming material diverted around the obstacles (between $t \approx 0.8\,\si{\second}$ and $1.0\,\si{\second}$), a negligible flow depth with a minimal velocity caused a sudden rise in $Fr_1$ before the material eventually coming to rest. 

The diverted material passing through the first row impacted the second row at $t \approx 0.45 \si{\second}$, resulting in a sudden increase in $Fr_2$. However, the material quickly deposited upstream of the second row, indicating the complete arrest of the flow. Because the material reaching the downstream region (Point 3) had already lost significant momentum, the Froude numbers ($Fr_3$) steadily approached zero, confirming that the flow eventually came to a complete rest.

The emergence of the dead zone and grain-free regions are consistent with previous experimental studies demonstrating that obstacle geometry critically governs flow interception \citep{Gray2003shockwaves, Cui2013gravity}. Ultimately, the formation of these regions confirms the applicability of the present numerical model, which successfully captures the underlying physical mechanisms by which obstacle height, width, and shape govern the overall flow dynamics of the array.

\subsection{Effect of obstacle height and width}
\label{subsec:obstacle_height_effect}

To analyze the effect of obstacle height (or width), nine cases were selected as shown in Table~\ref{tab:obstacle_cases}. The first three cases feature a uniform obstacle height across the array, while remaining six cases include height variations between rows in an array. All obstacles were cylindrical for the analysis presented in this subsection. To ensure a consistent comparison, the material volume of each obstacle was chosen to remain constant across all nine cases. For the range of obstacle heights evaluated in this section, the minimum height was selected based on the recommendation of \citet{ng2018dry}, which suggests that it should be at least three times the maximum flow depth observed in the absence of obstacles. Following \citet{yang2021numerical}, the obstacle width (or diameter) was chosen as twice the maximum flow depth without obstacles for the baseline case, resulting in an obstacle height of 0.06~\si{\meter} and a width of 0.048~\si{\meter}.
\begin{table}[width=.9\linewidth,cols=7,pos=h]
	\caption{Summary of uniform and variable obstacle array configurations, specifying the obstacle heights and widths across row 1 ($H_{o1}$ and $w_{o1}$), row 2 ($H_{o2}$ and $w_{o2}$), and row 3 ($H_{o3}$ and $w_{o3}$).}
	\label{tab:obstacle_cases}
	\begin{tabular*}{\tblwidth}{@{} LLLLLLL@{} }
		\toprule
		\textbf{Case} & \textbf{$H_{o1}$ (\si{\meter})} & \textbf{$w_{o1}$ (\si{\meter})} & \textbf{$H_{o2}$ (\si{\meter})} & \textbf{$w_{o2}$ (\si{\meter})} & \textbf{$H_{o3}$ (\si{\meter})} & \textbf{$w_{o3}$ (\si{\meter})}\\
		\midrule
		OH1 & 0.06 & 0.048 & 0.06 & 0.048 & 0.06 & 0.048 \\
		OH2 & 0.08 & 0.0416 & 0.08 & 0.0416 & 0.08 & 0.0416 \\
		OH3 & 0.10 & 0.0372 & 0.10 & 0.0372 & 0.10 & 0.0372 \\
		CS1 & 0.10 & 0.0372 & 0.08 & 0.0416 & 0.06 & 0.048 \\
		CS2 & 0.08 & 0.0416 & 0.10 & 0.0372 & 0.06 & 0.048 \\
		CS3 & 0.06 & 0.048 & 0.08 & 0.0416 & 0.10 & 0.0372 \\
		CS4 & 0.10 & 0.0372 & 0.06 & 0.048 & 0.08 & 0.0416 \\
		CS5 & 0.08 & 0.0416 & 0.06 & 0.048 & 0.10 & 0.0372 \\
		CS6 & 0.06 & 0.048 & 0.10 & 0.0372 & 0.08 & 0.0416 \\
		\bottomrule
	\end{tabular*}
\end{table}

All simulations were conducted on the synthetic curved surface previously mentioned in the validation study (Figure~\ref{fig:three_semisphere_obstacle}). To ensure consistency, the initial release volume, position, and friction parameters (internal and bed friction angles) were kept identical to the validation case. As the simulated avalanche descended the domain, it interacted with the arrays of cylindrical obstacles positioned along the flow path.

\begin{figure}[htbp]
	\begin{subfigure}[t]{0.48\textwidth}
		\centering
		\includegraphics[width=\textwidth]{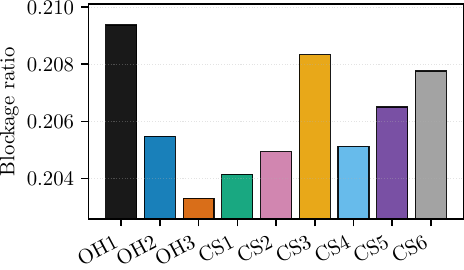}
		\caption{}
		\label{subfig:re_comparison_height}
	\end{subfigure}
	\hfill
	\begin{subfigure}[t]{0.48\textwidth}
		\centering
		\includegraphics[width=\textwidth]{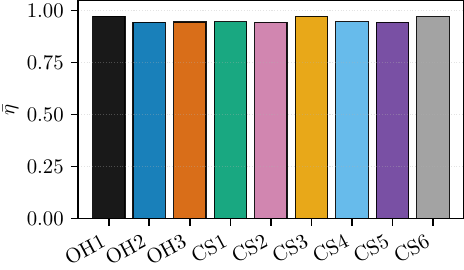}
		\caption{}
		\label{subfig:velocity_reduction_pt1_pt2}
	\end{subfigure}
	\caption{(a) Comparison of blockage ratio and (b) time-averaged velocity reduction ratio ($\bar{\eta}$) (between Point 1 and Point 2, marked in Figure~\ref{fig:array_cylinder}) for cases with uniform and variable obstacle heights. The uniform height cases are OH1, OH2, and OH3, while the variable height cases are denoted as CS1--CS6.}
	\label{fig:re_vrr_combined}
\end{figure}

To quantify the effectiveness of the array configurations, the blockage ratio (BR)---defined as the ratio of the initial release height to the final runout distance of the granular pile---was evaluated \citep{iverson2015entrainment}. A higher BR value indicates a greater reduction in flow mobility and a more efficient array. As illustrated in Figure~\ref{subfig:re_comparison_height}, the numerical results demonstrate that configurations OH1, CS3, and CS6 yielded the highest BR values across all cases. A comparative analysis of these three optimal cases reveals a common geometry feature: the leading row in each configuration features the minimum obstacle height chosen (i.e. 0.06~\si{\meter}). Because the material volume of each obstacle was kept constant throughout this study, minimizing the obstacle height proportionally maximizes its width, which significantly reduces the inter-obstacle spacing in the first row. Overall, these findings establish that under a constant volume constraint, the obstacle width is the primary deciding factor for maximizing the blockage ratio, a flow deceleration trend that aligns well with the findings of \citet{bi2019effects}.

To further evaluate the deceleration effect, the time-averaged velocity reduction ratio ($\bar{\eta}$) was adopted \citep{wang2017experimental} and measured between Point 1 and Point 2 (Figure~\ref{fig:array_cylinder}). The velocity reduction ratio is defined as:
\begin{equation}
    {\eta} = \frac{v_1 - v_2}{v_1}, 
\end{equation}
where $v_1$ and $v_2$ are the velocities at Point 1 and Point 2, respectively. While a higher $\bar{\eta}$ value theoretically signifies greater flow resistance, the numerical data revealed that $\bar{\eta}$ remained close to 1 across all tested configurations, as illustrated in Figure~\ref{subfig:velocity_reduction_pt1_pt2}. This uniformity occurred because the first row of the array significantly dissipated the initial momentum of the avalanche, abruptly arresting the flow and eliminating any noticeable variation in velocity between configurations, since the velocity at Point 2 was drastically lower than at Point 1. Because this massive initial deceleration yielded nearly identical velocity reduction ratios, array performance cannot be evaluated using this metric. Therefore, the blockage ratio serves as the only definitive selection criterion for determining the most effective array configuration in the present study.

\begin{figure}[htbp]
	\centering
	\begin{subfigure}[t]{0.48\textwidth}
		\centering
		\includegraphics[width=\linewidth]{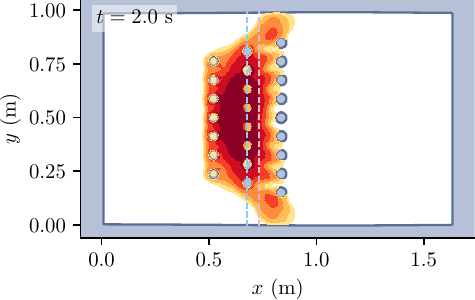}
	\end{subfigure}
	
	\vspace{0.4cm}

	\includegraphics{figures/height_contours_colorbar.pdf}
	
	\caption{Measurement of steady-state depth contours extent for OH1 configuration. The cylindrical obstacles are highlighted in gray, and grain-free region is shown in white. The blue and yellow vertical dashed lines denote centerline of second obstacle row and maximum extent of flow depth region, respectively.}
	\label{fig:deposition_depth_contour}
\end{figure}
\begin{figure}
    \centering
    \includegraphics{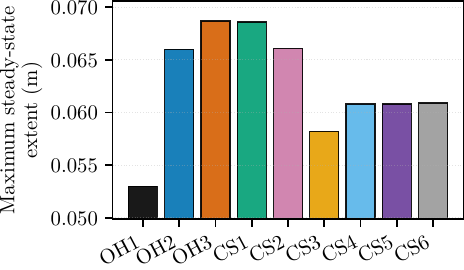}
    \caption{Comparison of maximum steady-state depth extent for uniform and variable obstacle height cases. The uniform height cases are OH1, OH2, and OH3, while the variable height cases are denoted as CS1--CS6.}
    \label{fig:maximum_steady_extend_comp}
\end{figure}

Drawing on the prior the BR analysis, configurations OH1 and CS3 clearly emerged as the most effective arrays. To identify the optimal case, the spatial extent of the maximum steady-state flow depth region (dark red) extending beyond the centerline of the second obstacle row was evaluated, as depicted in Figure~\ref{fig:deposition_depth_contour}. Fundamentally, the BR evaluates runout based on initial release conditions, whereas the maximum flow depth extent measures how far the material reaches past the second obstacle row. However, because the initial release height and position were held constant across all simulations in the present study, both metrics serve as similar indicators of flow mobility. Physically, a larger downstream extent beyond the second obstacle row signifies greater flow mobility and a consequently less efficient configuration. Upon comparing the OH1 and CS3 cases in Figure~\ref{fig:maximum_steady_extend_comp}, the OH1 array constrained this downstream extent to 0.0530~\si{\meter}, whereas the CS3 configuration allowed a nearly 10\% increase (an increment of 5.2~\si{\milli\meter}). This difference arises because the smaller inter-obstacle spacing in the OH1 configuration effectively arrests the granular mass before it can cross the second row, resulting in better material retention. These results firmly establish the OH1 case as the optimal configuration, justifying its selection for the subsequent investigation into obstacle shapes. Ultimately, across the BR and the maximum steady-state depth extent analyses, these findings collectively confirm that minimizing the inter-obstacle spacing in the first two rows of an array is the primary determining factor for maximizing overall array efficiency.

\subsection{Effect of obstacle shape}
\begin{figure}[htbp]
	\begin{subfigure}[t]{0.48\textwidth}
		\centering
		\includegraphics[width=\textwidth, height=0.8\textwidth, keepaspectratio]{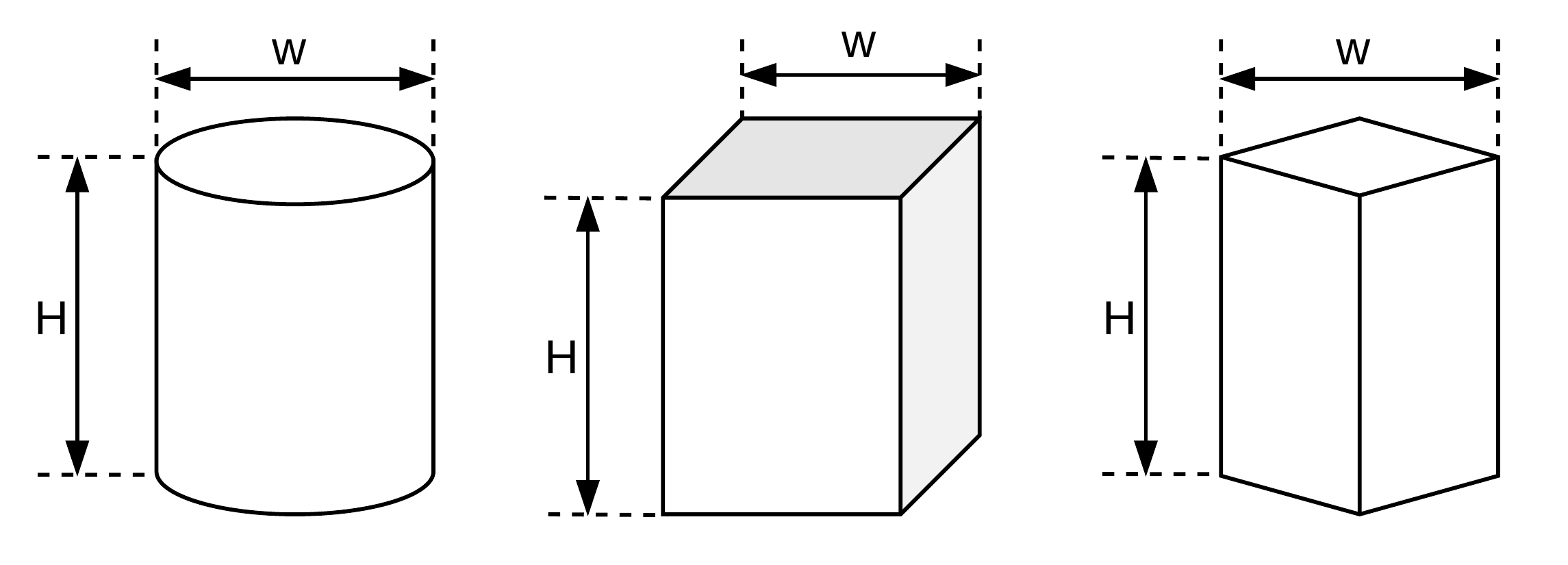}
		\caption{}
		\label{fig:individual_obstacle}
	\end{subfigure}
    \hfill
    \begin{subfigure}[t]{0.48\textwidth}
		\centering
		\includegraphics[width=\textwidth, height=0.8\textwidth, keepaspectratio]{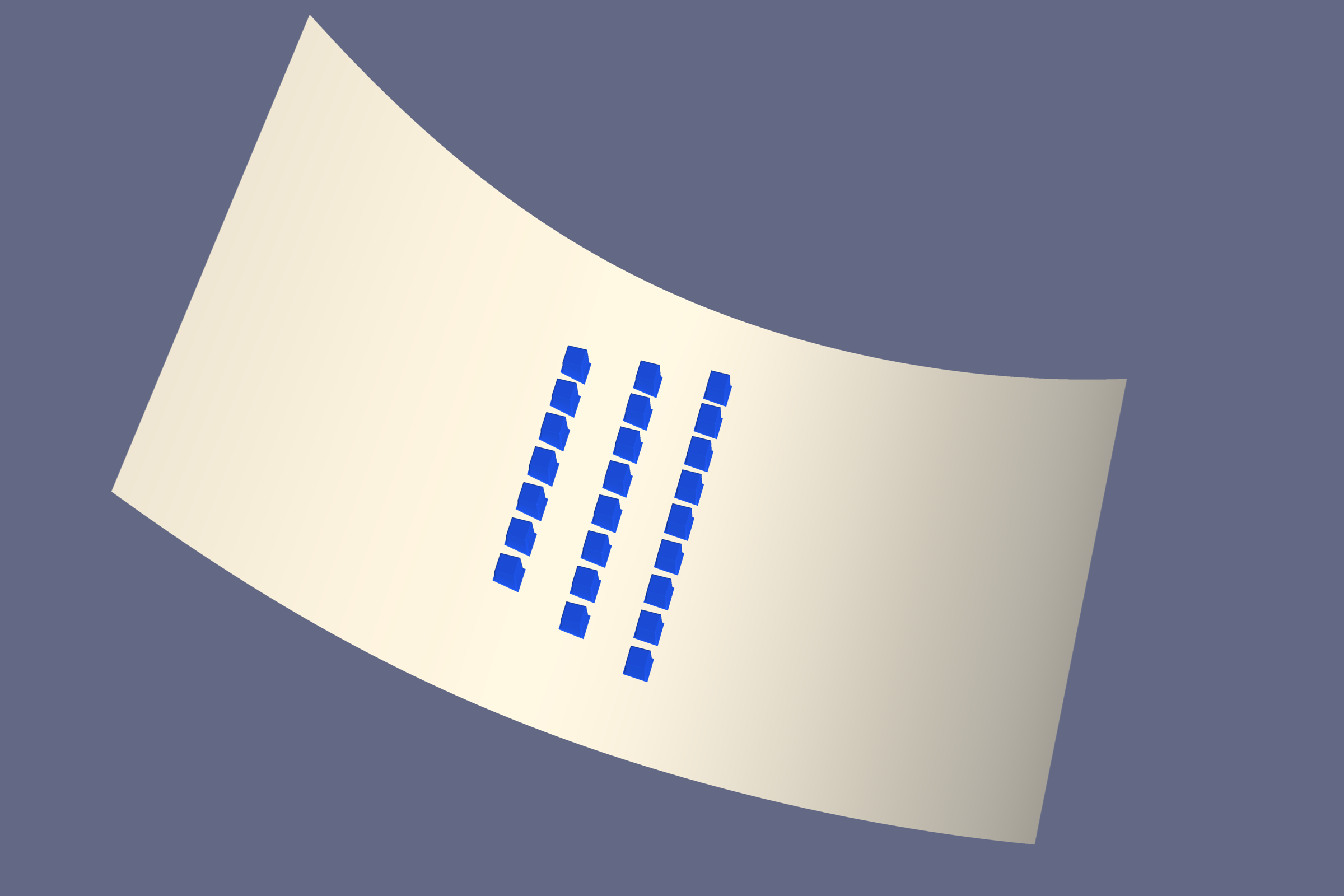}
		\caption{}
		\label{fig:array_cuboid}
	\end{subfigure}
	\caption{(a) Individual obstacle shapes: cylindrical (left), thin cuboid (center), and diamond cuboid (right) and (b) Array of uniform obstacle shape of thin cuboid.}
	\label{fig:array_of_same_shape}
\end{figure}

This section examines the effect of obstacle shape on flow properties including flow depth and velocity. In the present study, three obstacle shapes---cylindrical, thin cuboid, and diamond cuboid---were selected (Figure~\ref{fig:individual_obstacle}) based on their relevance in practical applications \citep{boensch2015technical,mcclung2006avalanche}. The results were evaluated using the maximum flow accumulation upstream of each obstacle in an array, blockage ratio, and maximum steady-state depth extent to identify the optimal array shape configuration.

In total, nine array configurations were evaluated, comprising three uniform and six mixed-shape arrangements (Table~\ref{tab:obstacle_shape_cases}). While all obstacles maintained a constant height ($H$) of $0.06~\si{\meter}$, their width ($w$) varied according to the specific geometry, as illustrated in Figure~\ref{fig:individual_obstacle}. In the uniform configurations, all three rows consisted of a single obstacle shape (Figure~\ref{fig:array_cuboid}), whereas each row featured a distinct shape in the mixed configurations.

\begin{table}[width=.9\linewidth,cols=4,pos=h]
	\caption{Nomenclature and row-wise distribution of uniform and mixed-shape obstacle configurations.}
	\label{tab:obstacle_shape_cases}
	\begin{tabular*}{\tblwidth}{@{} LLLL@{} }
		\toprule
		\textbf{Configuration name} & \textbf{First row} & \textbf{Second row} & \textbf{Third row} \\
		\midrule
		AC & Cylindrical & Cylindrical & Cylindrical \\
		ACu & Thin cuboid & Thin cuboid & Thin cuboid \\
		ADC & Diamond cuboid & Diamond cuboid & Diamond cuboid \\
		CDT & Cylindrical & Diamond cuboid & Thin cuboid \\
		CTD & Cylindrical & Thin cuboid & Diamond cuboid \\
		DCT & Diamond cuboid & Cylindrical & Thin cuboid \\
		DTC & Diamond cuboid & Thin cuboid & Cylindrical \\
		TCD & Thin cuboid & Cylindrical & Diamond cuboid \\
		TDC & Thin cuboid & Diamond cuboid & Cylindrical \\
		\bottomrule
	\end{tabular*}
\end{table}

\begin{figure}[htbp]
	\centering
	
	\begin{subfigure}[t]{0.31\textwidth}
		\centering
		\includegraphics[width=\linewidth]{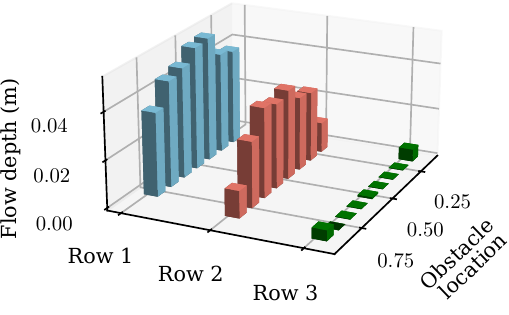}
		\caption{ADC}
	\end{subfigure}\hfill
	\begin{subfigure}[t]{0.31\textwidth}
		\centering
		\includegraphics[width=\linewidth]{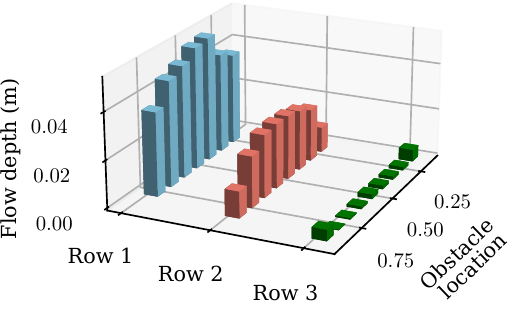}
		\caption{DCT}
	\end{subfigure}\hfill
	\begin{subfigure}[t]{0.31\textwidth}
		\centering
		\includegraphics[width=\linewidth]{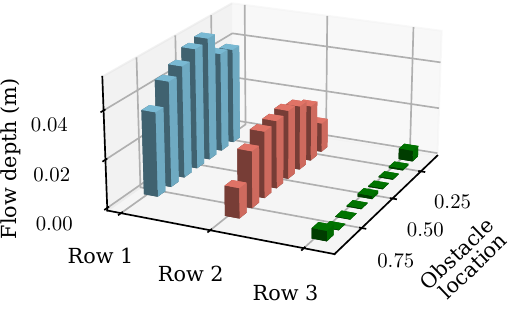}
		\caption{DTC}
	\end{subfigure}
	
	\vspace{0.4cm}
	
	\begin{subfigure}[t]{0.31\textwidth}
		\centering
		\includegraphics[width=\linewidth]{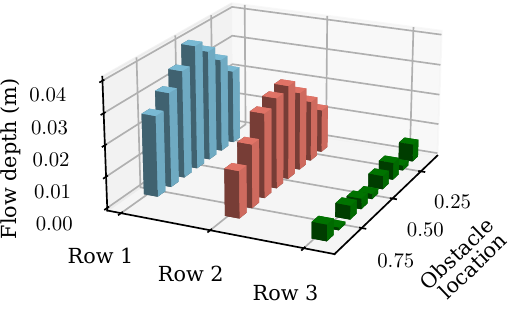}
		\caption{TCD}
	\end{subfigure}\hfill
	\begin{subfigure}[t]{0.31\textwidth}
		\centering
		\includegraphics[width=\linewidth]{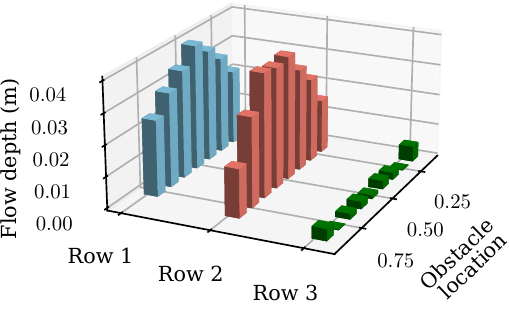}
		\caption{TDC}
	\end{subfigure}\hfill
	\begin{subfigure}[t]{0.31\textwidth}
		\centering
		\includegraphics[width=\linewidth]{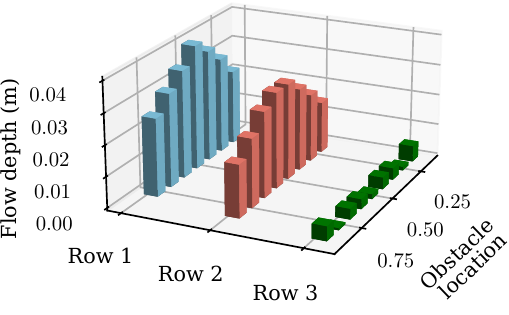}
		\caption{ACu}
	\end{subfigure}
	
	\vspace{0.4cm}
	
	\begin{subfigure}[t]{0.31\textwidth}
		\centering
		\includegraphics[width=\linewidth]{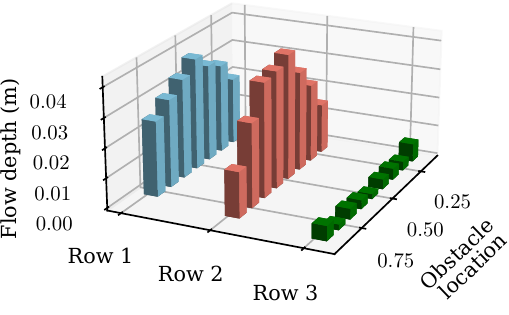}
		\caption{CDT}
	\end{subfigure}\hfill
	\begin{subfigure}[t]{0.31\textwidth}
		\centering
		\includegraphics[width=\linewidth]{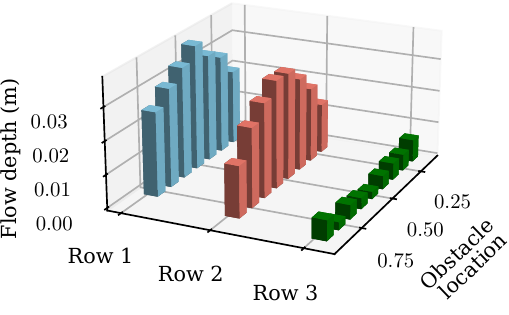}
		\caption{CTD}
	\end{subfigure}\hfill
	\begin{subfigure}[t]{0.31\textwidth}
		\centering
		\includegraphics[width=\linewidth]{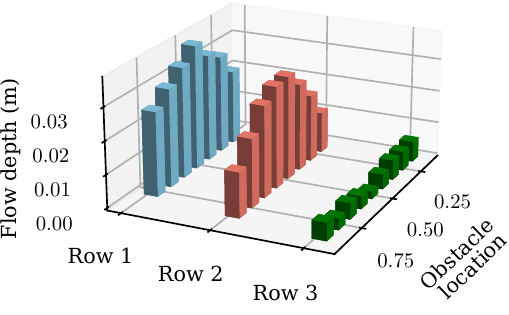}
		\caption{AC}
	\end{subfigure}
	\caption{Flow depth variation upstream of each obstacle across different rows in arrays with uniform and mixed-shape obstacles: (a) ADC, (b) DCT, (c) DTC, (d) TCD, (e) TDC, (f) ACu, (g) CDT, (h) CTD, and (i) AC. The blue, red, and green bars represent the maximum flow depth upstream of obstacles in the first, second, and third rows, respectively.}
	\label{fig:pile_height_histogram_all}
\end{figure}

To understand the physical mechanism behind the overall effectiveness of the obstacle shape, upstream flow deposition was examined across the three rows of an array (Figure~\ref{fig:pile_height_histogram_all}). Configurations featuring diamond cuboid obstacles in the first row (ADC, DCT, and DTC) were the most effective at resisting the avalanche flow, recording the highest maximum flow depth upstream of the first row. This accumulation is driven by the formation of a granular hydraulic jump. The diamond geometry abruptly decelerates the incoming flow, forcing a transition into a slower-moving flow regime. This jump induces rapid kinetic energy dissipation, severely blocking the granular material from passing through the obstacle gaps. Consequently, this strong restriction drastically limits the material reaching subsequent rows, yielding significantly lower upstream flow depths in the second and third rows compared to the remaining configurations. In contrast, cylindrical obstacles in the first row (CDT, CTD, and AC) allowed material to pass through the obstacle gaps easily, resulting in comparatively higher deposition upstream of the second and third rows relative to arrays with diamond and thin cuboid shapes in the first row. A notable exception is the TDC case, where the diamond cuboid obstacle in the second row entrapped the flow, resulting in a greater upstream flow depth. Overall, these observations establish that the diamond geometry acts as a severe flow obstruction, and locating it in the first row effectively intercepts the initial flow as well as severely blocks downstream material propagation.

\begin{figure}[htbp]
		\centering
		\includegraphics{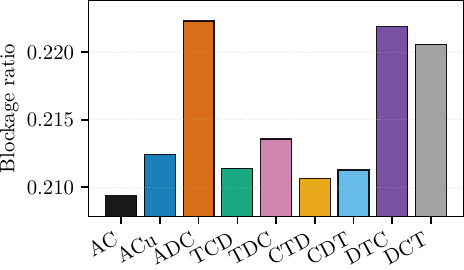}
	\caption{Comparison of blockage efficiency for uniform and mixed-shape obstacle arrays. The uniform shape cases are AC, ACu, and ADC, while the mixed-shape cases are denoted as TCD, TDC, CTD, CDT, DTC, and DCT.}
	\label{fig:runout_eff_comp}
\end{figure}

To further assess the overall effectiveness of the array configurations based on avalanche runout distance, the blockage ratio for the shape configurations was evaluated. As discussed in Section~\ref{subsec:obstacle_height_effect}, BR quantifies the granular material's mobility. Figure~\ref{fig:runout_eff_comp} depicts that leading diamond cuboid obstacles consistently produced the most substantial blockage ratio. The uniform ADC configuration exhibited the highest BR value, closely followed by the mixed DTC and DCT configurations. Consistent with the flow depth analyses, the severe initial deposition caused by the leading diamond obstacles effectively restricted downstream flow mobility, yielding significantly shorter runout distances and further highlighting the impact of the granular hydraulic jump. These results demonstrate that placing the diamond-shaped obstacles in the first row is a highly effective structural strategy for minimizing material flow propagation.

\begin{figure}
    \centering
    \includegraphics{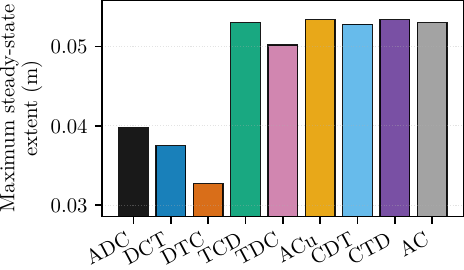}
    \caption{Comparison of maximum steady-state depth extent for uniform and mixed-shape obstacle arrays. The uniform shape cases are AC, ACu, and ADC, while the mixed-shape cases are denoted as TCD, TDC, CTD, CDT, DTC, and DCT.}
    \label{fig:maximum_steady_extend_mixed_comp}
\end{figure}

In addition to the prior analysis of maximum flow depth and BR, the spatial extent of the maximum steady-state flow depth region was evaluated, as discussed in Section~\ref{subsec:obstacle_height_effect}. Specifically, to observe the effect of the leading row obstacle geometry, the downstream extent beyond the second row was compared between the DTC and AC configurations. As illustrated in Figure~\ref{fig:maximum_steady_extend_mixed_comp}, the DTC case exhibited the minimum downstream extent of 0.0398 \si{\meter}. In contrast, the AC array demonstrated a nearly 30\% increase in downstream extent relative to the DTC configuration. Ultimately, this spatial steady-state analysis quantitatively supports the earlier findings, confirming that the DTC configuration---featuring diamond-shaped obstacles in the first row---is the most effective at entrapping the granular material.

In practical engineering, the design of protective obstacle systems requires careful consideration of the peak impact force acting on the structures. The maximum impact force exerted by the granular material is a critical parameter, as it dictates the required damage resistance of the structures and directly influences construction costs. 

The impact force exerted by the granular flow is estimated from the rate of change of momentum during the interaction. The final expression is given as:
\begin{equation}
	F_{\text{impact}} = \frac{1}{2} \rho_f g_z\, hA + \rho_f A \bar{u}^2,
\end{equation}
where $\rho_f$ is the effective flow density of the granular material. Here, the first term on the right hand side represents the hydrostatic contribution, while the second term accounts for the dynamic impact driven by incoming flow momentum. Since the granular mass did not overtop the obstacles, the effective impact area was defined using the projected obstacle width ($w$) as $A = w h$. Using the upstream flow depth ($h$) and depth-averaged velocity ($\bar{u}$), the spatial impact force field was computed within a zone extending $3~\si{\centi\meter}$ upstream of each row and spanning laterally across the overall width of the respective row. From this upstream force field, the peak value was extracted as the maximum impact force, serving as a representative design load for the obstacles within that row to withstand avalanche impact.

\begin{figure}[htbp]
	\centering
	
	\begin{subfigure}[t]{0.31\textwidth}
		\centering
		\includegraphics[width=\linewidth]{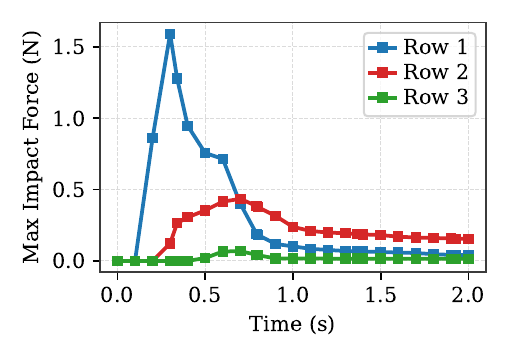}
		\caption{ADC}
	\end{subfigure}\hfill
	\begin{subfigure}[t]{0.31\textwidth}
		\centering
		\includegraphics[width=\linewidth]{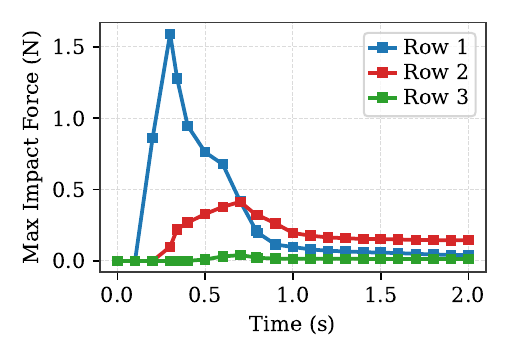}
		\caption{DCT}
	\end{subfigure}\hfill
	\begin{subfigure}[t]{0.31\textwidth}
		\centering
		\includegraphics[width=\linewidth]{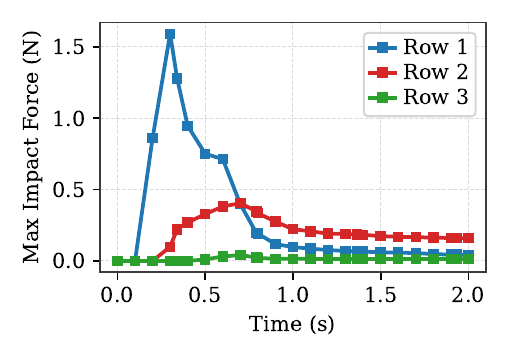}
		\caption{DTC}
	\end{subfigure}
	
	\vspace{0.4cm}
	
	\begin{subfigure}[t]{0.31\textwidth}
		\centering
		\includegraphics[width=\linewidth]{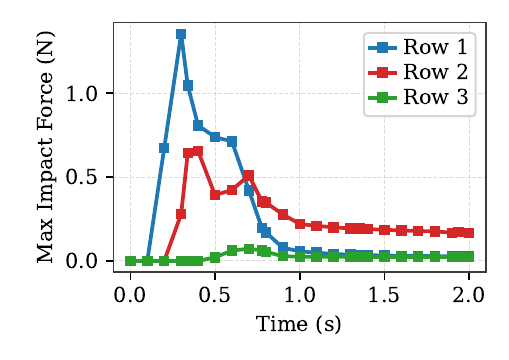}
		\caption{TCD}
	\end{subfigure}\hfill
	\begin{subfigure}[t]{0.31\textwidth}
		\centering
		\includegraphics[width=\linewidth]{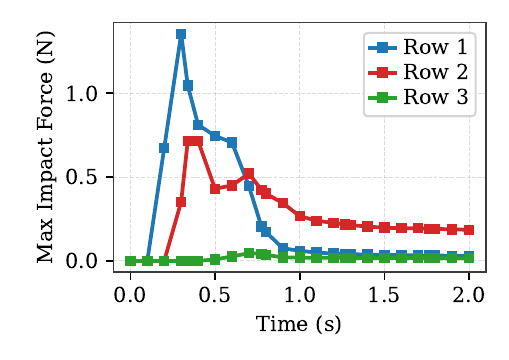}
		\caption{TDC}
	\end{subfigure}\hfill
	\begin{subfigure}[t]{0.31\textwidth}
		\centering
		\includegraphics[width=\linewidth]{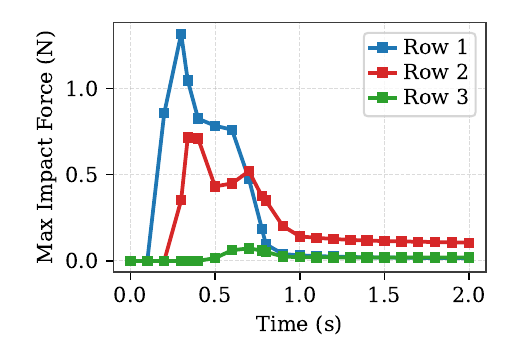}
		\caption{ACu}
	\end{subfigure}
	
	\vspace{0.4cm}
	
	\begin{subfigure}[t]{0.31\textwidth}
		\centering
		\includegraphics[width=\linewidth]{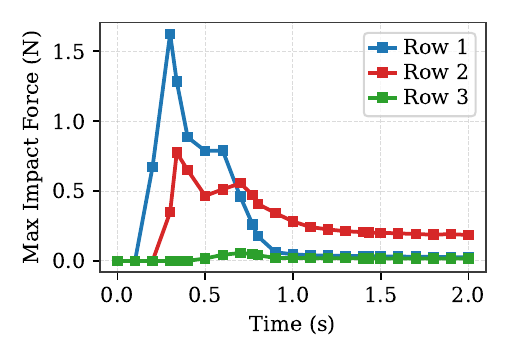}
		\caption{CDT}
	\end{subfigure}\hfill
	\begin{subfigure}[t]{0.31\textwidth}
		\centering
		\includegraphics[width=\linewidth]{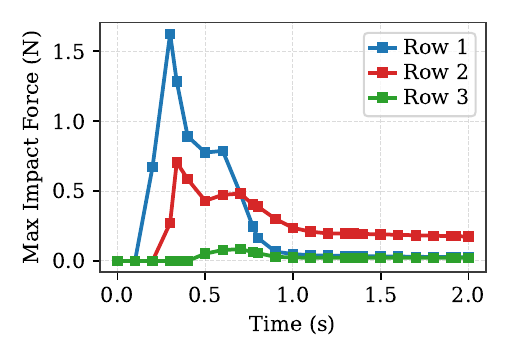}
		\caption{CTD}
	\end{subfigure}\hfill
	\begin{subfigure}[t]{0.31\textwidth}
		\centering
		\includegraphics[width=\linewidth]{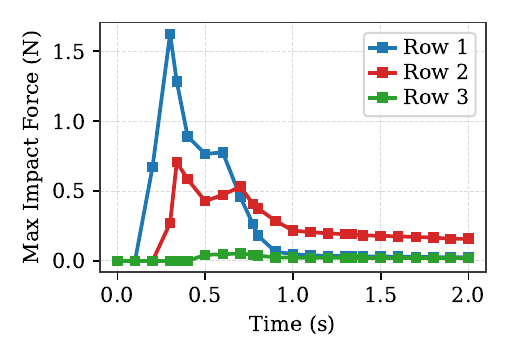}
		\caption{AC}
	\end{subfigure}
	
	\caption{Comparison of maximum impact force upstream of each obstacle across different rows for arrays with uniform and mixed-shape obstacle configuration. (a) ADC, (b) DCT, (c) DTC, (d) TCD, (e) TDC, (f) ACu, (g) CDT, (h) CTD, and (i) AC. The effective flow density of the granular material is taken approximately as 1550~\si{\kilogram\per\cubic\meter}, representative of silica sand, and the obstacle width is 0.048~\si{\meter}.}
	\label{fig:max_impact_force}
\end{figure}

Figure~\ref{fig:max_impact_force} presents the variation of impact force over time. For these calculations, an effective flow density of 1550~\si{\kilogram\per\cubic\meter} was adopted for silica sand, corresponding to a solid density of 2650~\si{\kilogram\per\cubic\meter} \citep{li2022insight}, which aligns closely with experimental flow densities reported for glass beads ($1380~\si{\kilogram\per\cubic\meter}$; \citealt{hauksson2007laboratory}). In general, the particle density of silica sand is higher than that of glass beads, hence a higher value of flow density in the calculation of impact force is considered. Further, the obstacle width (or projected width for cylindrical and diamond shapes) were used. The results reveal that diamond cuboid obstacles positioned in the first row experienced the maximum impact force, followed by the thin cuboid and cylindrical obstacles in the first row. As these leading diamond cuboids absorbed peak initial impact, the maximum force exerted on the second row was significantly reduced compared to all other configurations. Consistent with earlier studies \citep{bi2018numerical, bi2019effects}, the impact force was consistently higher on the first row than on  subsequent rows. These force distributions provide further evidence that placing diamond cuboid obstacles in the leading row is a highly effective strategy for blocking and dissipating incoming avalanche energy.

\subsection{Empirical correlation}
In the presence of an obstacle array, accurately predicting the maximum flow depth, its spatial location, and its mean flow velocity corresponding to location of maximum flow depth is crucial for designing effective avalanche mitigation strategies. To enable the direct estimation of these parameters without requiring computationally intensive simulations, an empirical correlation for unsteady and non-uniform granular flows is proposed. This correlation was derived by extracting the key flow variables from the simulation data following avalanche-obstacle interaction across all array configurations, including uniform heights, variable heights, and uniform and mixed-shape configurations (See Appendix~\ref{app:empirical}). A comprehensive comparison of these numerical results revealed a mathematical relation between the flow characteristics and the specific obstacle geometries (height, width, and shape).

\subsubsection{Maximum flow depth correlation}\label{subsubsec:hmax_corelation}
To explicitly quantify this relationship, the best-fit correlation derived from the numerical data for the maximum flow depth is expressed as:
\begin{equation}
	h_{\text{max}} = C\, h_{0}\, \left(\frac{w_1}{h_0}\right)^{\eta} \left(\frac{H_1}{h_0}\right)^{\kappa} \exp{\left(\beta\, \text{sgn}(\Delta H) + \gamma\, \text{sgn}^2 (\Delta H)\right)},
	\label{eq:corelation_h_max}
\end{equation}
where $h_{\text{max}}$ is the maximum flow depth during avalanche-obstacle interaction, $C$ is the shape factor, and $h_0$ is the initial release height. The parameters $w_1$ and $H_1$ denote the width and height of the obstacles in the first row, respectively. The exponents $\eta$, $\kappa$, $\beta$, and $\gamma$ are empirical constants that depend on the specific array configuration. The term $\Delta H$ represents a dimensionless obstacle height gradient, defined as:
\begin{equation}
 \Delta H= \frac{H_3 - 1.1H_1 + 0.009}{h_0}.  
\end{equation}
Finally, $\text{sgn}$ denotes the signum function.

The correlation presented in Equation~\ref{eq:corelation_h_max} indicates that for most array configurations except diamond-leading, the maximum flow depth is significantly influenced by the height and width of the first-row obstacles, highlighted in Section~\ref{subsec:obstacle_height_effect}. As detailed in Table~\ref{tab:corelation_contant_hmax}, the exponents $\eta$ and $\kappa$ remain constant across all configurations (except diamond shape) with different leading-obstacle shapes, with the flow variations captured entirely by changes in the shape factor. In contrast, for diamond-leading configuration, the influence of the leading obstacle dimensions becomes negligible ($\eta = \kappa = 0$). Instead, the maximum flow depth in the diamond-leading arrangement is governed exclusively by the obstacle height gradient across the array, dictated by the parameters $\beta$ and $\gamma$.

\begin{table}[width=.9\linewidth,cols=6,pos=h]
	\caption{Summary of empirical constants for the maximum flow depth correlation, with a corresponding correlation coefficient ($R$) of 0.953176}
	\label{tab:corelation_contant_hmax}
	\begin{tabular*}{\tblwidth}{@{} LLLLLL@{} }
		\toprule
		\textbf{Obstacle configurations} & \textbf{C} & $\boldsymbol{\eta}$ & $\boldsymbol{\kappa}$ & $\boldsymbol{\beta}$ & $\boldsymbol{\gamma}$\\
		\midrule
		Cylinder-leading & 0.4657 & 0.002 & 0.033 & 0 & 0\\
		Thin cuboid-leading & 0.4946 & 0.002 & 0.033 & 0 & 0\\
		Diamond-leading & 0.46 & 0 & 0 & 0.0029 & 0.2369 \\
		\bottomrule
	\end{tabular*}
\end{table}

\subsubsection{Spatial location of maximum flow depth}
While Equation~\ref{eq:corelation_h_max} quantifies the magnitude of the maximum flow depth, determining its exact spatial location is equally important for understanding deposition behavior. Accordingly, empirical correlations for these spatial coordinates were derived from the numerical data and are expressed as:
\begin{equation}
	X_{\text{cen}} = C\, h_{0}\, \left(\frac{A_{\text{tot}}}{{h_0}^2}\right)^{\alpha}  \left(\frac{w_1}{h_0}\right)^{\eta} \left(\frac{H_1}{h_0}\right)^{\kappa} \exp{\left(\beta\, \text{sgn}(\Delta H) + \gamma\, \text{sgn}^2(\Delta H)\right)},
	\label{eq:corelation_xcen}
\end{equation}
and
\begin{equation}
	Y_{\text{cen}} = C\, h_{0}\, \left(\frac{A_{\text{tot}}}{{h_0}^2}\right)^{\alpha}  \left(\frac{w_1}{h_0}\right)^{\eta} \left(\frac{H_1}{h_0}\right)^{\kappa} \exp{\left(\beta\, \text{sgn}(\Delta H) + \gamma\, \text{sgn}^2(\Delta H)\right)},
	\label{eq:corelation_ycen}
\end{equation}
where $X_{\text{cen}}$ and $Y_{\text{cen}}$ denote the spatial coordinates corresponding to the maximum flow depth. The term $A_{\text{tot}}$ represents the cumulative area of the obstacles across the array, defined as $A_{\text{tot}} = n_1H_1 w_1 + n_2H_2 w_2 + n_3H_3 w_3$, where $n_1, n_2$ and $n_3$ denote number of obstacles present in the corresponding rows. The remaining parameters retain the definitions established in Section~\ref{subsubsec:hmax_corelation}.

Equations~\ref{eq:corelation_xcen} and \ref{eq:corelation_ycen} reveal distinct behaviors for the longitudinal ($X_{\text{cen}}$) and lateral ($Y_{\text{cen}}$) coordinates of the maximum flow depth. As detailed in Table~\ref{tab:corelation_contant_xcen}, the longitudinal position $X_{\text{cen}}$ for configurations with varying leading shapes depends not only on the height and width of the first row but also on the cumulative obstacle area. Conversely, Table~\ref{tab:corelation_contant_ycen} demonstrates that the lateral position $Y_{\text{cen}}$ remains virtually influenced by obstacle geometry only. The near-zero exponent values across all configurations illustrate minimal lateral deviation in the simulation data, indicating that $Y_{\text{cen}}$ remains largely unaffected by variations in obstacle height and width, with obstacle shape exerting only a marginal influence.

\begin{table}[width=.9\linewidth,cols=7,pos=h]
	\caption{Empirical constants for the $X_{\text{cen}}$ correlation, with a corresponding R value of 0.99767}
	\label{tab:corelation_contant_xcen}
	\begin{tabular*}{\tblwidth}{@{} LLLLLLL@{} }
		\toprule
		\textbf{Obstacle configurations} & \textbf{C} & $\boldsymbol{\alpha}$ & $\boldsymbol{\eta}$ & $\boldsymbol{\kappa}$ & $\boldsymbol{\beta}$ & $\boldsymbol{\gamma}$ \\
		\midrule
		Cylinder-leading & 3.0756 & -0.0004 & 0.0076 & 0.0143 & 0.0001 & 0.0001 \\
		Thin cuboid-leading & 2.9221 & 0.8997 & 1.3354 & 0.4298 & -0.0058 & -0.0053 \\
		Diamond-leading & 4.0277 & 0.0133 & -0.0764 & -0.0064 & -0.0606 & 0.0009 \\
		\bottomrule
	\end{tabular*}
\end{table}

\begin{table}[width=.9\linewidth,cols=7,pos=h]
	\caption{Empirical constants for the $Y_{\text{cen}}$ correlation, with a corresponding R value of 0.966944}
	\label{tab:corelation_contant_ycen}
	\begin{tabular*}{\tblwidth}{@{} LLLLLLL@{} }
		\toprule
		\textbf{Obstacle configurations} & \textbf{C} & $\boldsymbol{\alpha}$ & $\boldsymbol{\eta}$ & $\boldsymbol{\kappa}$ & $\boldsymbol{\beta}$ & $\boldsymbol{\gamma}$ \\
		\midrule
		Cylinder-leading & 4.2371 & 0 & 0 & 0 & 0 & 0 \\
		Thin cuboid-leading & 4.2372 & -0.0008 & -0.0012 & -0.0004 & 0 & 0 \\
		Diamond-leading & 4.2354 & 0.0003 & 0.0005 & 0.0003 & 0.0001 & 0 \\
		\bottomrule
	\end{tabular*}
\end{table}

\subsubsection{Mean velocity correlation}
To complete the kinematic characterization of the granular flow, an empirical correlation for the mean velocity ($V_{\text{mean}}$) at the location of maximum flow depth was developed. Derived directly from the numerical data, this relationship is expressed as:
\begin{equation}
	V_{\text{mean}} = C\, \sqrt{gh_{0}}\, \left(\frac{A_{\text{tot}}}{{h_0}^2}\right)^{\alpha} \left(\frac{w_1}{h_0}\right)^{\eta} \left(\frac{H_1}{h_0}\right)^{\kappa} \exp{\left(\beta\, \text{sgn}(\Delta H) + \gamma\, \text{sgn}^2(\Delta H)\right)}.
	\label{eq:corelation_vmean}
\end{equation}
As detailed in Table~\ref{tab:corelation_contant_vmean}, Equation~\ref{eq:corelation_vmean} reveals that for configurations with varying leading shapes, the mean velocity is influenced by the cumulative obstacle area, height and width of the first-row obstacles. Additionally, the mean velocity for the diamond-leading arrangement is highly influenced by the obstacle height gradient of the array in comparison to the remaining cases. Overall, these mathematical correlations provide a comprehensive, computationally efficient toolkit for predicting critical avalanche--obstacle interaction parameters without the need for extensive numerical simulations.

\begin{table}[width=.9\linewidth,cols=5,pos=h]
	\caption{Correlation coefficients and fitted constants for $V_{\text{mean}}$, with a corresponding R value of 0.999749}
	\label{tab:corelation_contant_vmean}
	\begin{tabular*}{\tblwidth}{@{} LLLLLLL@{} }
		\toprule
		\textbf{Obstacle configurations} & \textbf{C} & $\boldsymbol{\alpha}$ & $\boldsymbol{\eta}$ & $\boldsymbol{\kappa}$ & $\boldsymbol{\beta}$ & $\boldsymbol{\gamma}$ \\
		\midrule
		Cylinder-leading & 0.9397 & 0.0058 & -0.0187 & 0.0242 & -0.0004 & 0.0002 \\
		Thin cuboid-leading & 0.9430 & -0.0456 & -0.0771 & -0.0445 & -0.0035 & -0.0022 \\
		Diamond-leading & 0.6754 & 0.0714 & 0.1719 & 0.1924 & 0.1598 & -0.0072 \\
		\bottomrule
	\end{tabular*}
\end{table}

\subsection{Application to field-scale avalanche} \label{sec6}
\begin{figure}[h!]
	\centering
	\includegraphics[width=1.0\textwidth]{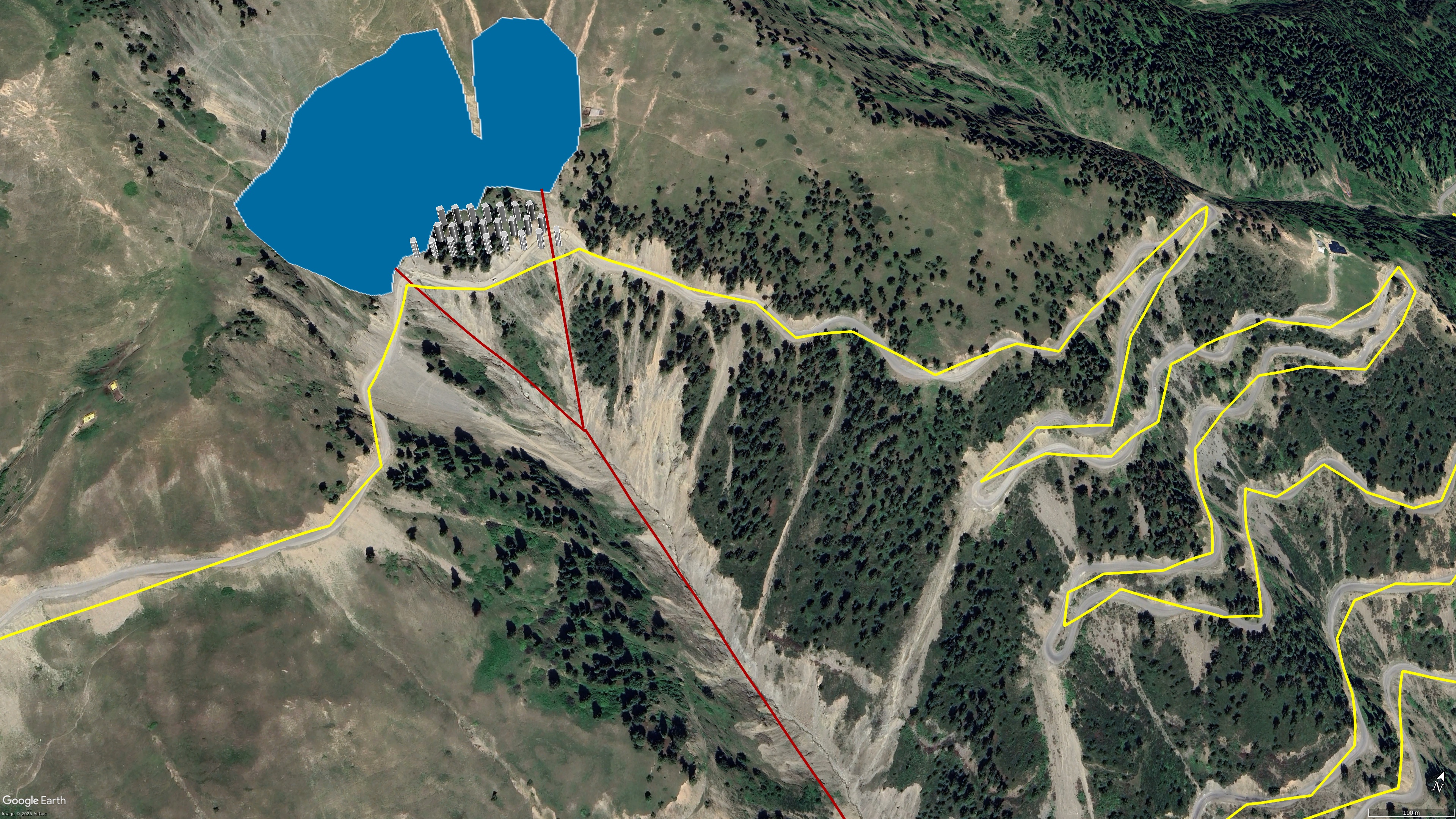}
	\caption{The CT-8 avalanche site with arrays of mixed obstacles inserted into it. The blue area shows starting position of avalanche, red lines represents avalanche path, and the mixed-shape obstacle arrays are represented by white pillars. \emph{Courtesy: Google Earth satellite imagery.}}
	\label{fig:ct_site_obstacles}
\end{figure}

Located in the Indian Western Himalayas, the Chowkibal-Tangdhar road in the union territory of Jammu and Kashmir is highly susceptible to dry snow avalanches, with 26 distinct active sites identified by the Defence Geoinformatics Research Establishment (DGRE) \citep{singh2020simulation}. The present study focuses on the CT-8 avalanche site (34$^\circ$24'58" N, 73$^\circ$57'48" E), situated approximately 6 \si{\kilo\meter} from Chowkibal village toward Tangdhar. As illustrated in Figure~\ref{fig:ct_site_obstacles}, the site is characterized by sparse tree cover and two primary flow paths, identified as the left and right gullies. Topographic analysis reveals that the south-to-southeast facing release zone lies at an elevation of 3000--3500 m a.s.l., with upper terrain slope angles ranging from 30$^\circ$ to 45$^\circ$ and the lower runout terrain flattening to between 12$^\circ$ and 25$^\circ$. The destructive potential of this specific site was demonstrated on January 5, 2018, when a severe avalanche swept a vehicle into a deep valley, resulting in ten fatalities and two rescues.

In the present study, the historical avalanche at the CT-8 site was simulated using the \texttt{TITAN2D} code, as for the previous cases. Using the DEM, release zone area, fracture height, and friction parameters introduced earlier, the solver performs the subsequent simulations. The actual topography was incorporated using a 12.5~\si{\meter} resolution DEM obtained from the Alaska Satellite Facility. Following \citet{singh2020simulation}, the total release area was defined as approximately 61,800~\si{\square\meter} with an initial fracture height of 25~\si{\centi\meter}. The numerical solver has been shown to accurately model dry (cohesionless) avalanches using a Mohr--Coulomb rheology. Accordingly, an internal friction angle ($\phi$) of 15$^\circ$ and a bed friction angle ($\delta$) of 27.5$^\circ$ were obtained. These specific values were selected based on the excellent agreement with the field data presented by \citet{sharma2026computational}. Given the demonstrated capability of the numerical solver, obstacle arrays were subsequently introduced along the CT-8 flow path in the present study to examine their influence on flow behavior and deposition characteristics.

\begin{figure}[htbp]
	\centering
	\begin{subfigure}[t]{0.48\textwidth}
		\centering
		\includegraphics{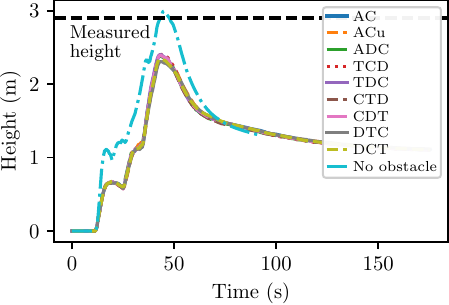}
		\caption{}
		\label{fig:ct_8_height_comparison}
	\end{subfigure}
	\begin{subfigure}[t]{0.48\textwidth}
		\centering
		\includegraphics{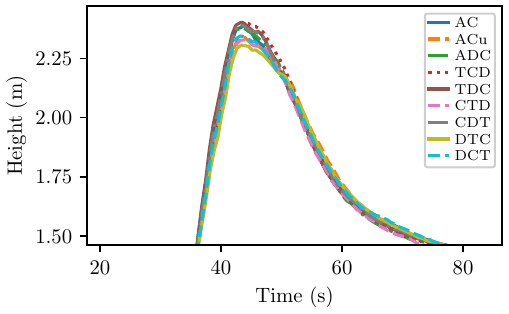}
		\caption{}
		\label{fig:ct_8_magnified_height}
	\end{subfigure}
	\caption{Flow depth variation during snow avalanche interaction with uniform and mixed-shape obstacle array configurations, along with the no obstacle case, near the roadside region at the CT-8 site. Figure~\ref{fig:ct_8_height_comparison} presents the temporal evolution of flow depth with a comparison to the actual data. Figure~\ref{fig:ct_8_magnified_height} shows a magnified view of the flow depth variations for uniform and mixed-shape obstacle configurations. The simulation input parameters are: release area = 61,800 \si{\square\meter}, release height = 25 \si{\centi\meter}, internal friction angle $\phi = 15^\circ$, and bed friction angle $\delta = 27.5^\circ$.}
\end{figure}

Figure~\ref{fig:ct_site_obstacles} illustrates the three-row obstacle array configurations overlaid on the actual CT-8 avalanche terrain, with each obstacle measuring 9 \si{\meter} in height and 6 \si{\meter} in width. Following the optimal layout criteria established by \citet{bi2022optimal}, the arrays were configured with an inter-obstacle spacing ($S_c$) of 6.3 \si{\meter} and an inter-row spacing ($S_r$) of 3.5 \si{\meter}. To intercept the bulk of the moving avalanche, the arrays were positioned directly within the flow path, with the central obstacle of the first-row placed at 34$^\circ$24'38.48" N, 73$^\circ$57'52.33" E. Using this obstacle-terrain layout, numerical simulations were conducted across both uniform and mixed-shape configurations to evaluate their influence on flow depth and mean velocity. Figure~\ref{fig:ct_8_height_comparison} presents the resulting temporal variation in flow depth near the roadside region---marked by the yellow lines in Figure~\ref{fig:ct_site_obstacles}---comparing the various obstacle configurations against the no-obstacle baseline case.

The obstacle arrays interacted with the avalanche, before it reached the road, disrupting the flow and trapping a portion of the material within the structures. Consequently, a reduced volume of material passed through the roadside region, as clearly visible in Figure~\ref{fig:ct_8_height_comparison}. Compared to the no-obstacle baseline, the presence of the arrays reduced the flow depth near the road by approximately 20\%.

To highlight the specific effects of the varying obstacle geometries, a magnified view of the flow depth variation is presented in Figure~\ref{fig:ct_8_magnified_height}. Table~\ref{tab:obstacle_cases_CT8} illustrates the percentage change in flow depth for the various obstacle configurations relative to the no-obstacle case. Among all evaluated configurations, the DTC configuration yielded the lowest flow depth near the road. However, the variations between the different obstacle shapes remained marginal. While the simulation results on idealized curved surface demonstrated a pronounced sensitivity to obstacle shapes, the complexity of the real-world terrain appears to dominate the flow dynamics, thereby diminishing the influence of individual obstacle shapes.

\begin{table}[width=.9\linewidth,cols=3,pos=h]
    \caption{Comparison of the effect of uniform and mixed-shape obstacle configurations on flow depth relative to the no-obstacle baseline for the real-world CT-8 avalanche site.}
    \label{tab:obstacle_cases_CT8}
    \begin{tabular*}{\tblwidth}{@{} LLL@{} }
        \toprule
        \textbf{Configuration} &
        \makecell[l]{\textbf{Maximum flow}\\
                     \textbf{depth (\si{\meter})}} &
        \makecell[l]{\textbf{Reduction in maximum}\\
                     \textbf{flow depth (\%)}} \\
        \midrule
        AC  & 2.38 & 20.67 \\
        ACu & 2.31 & 23 \\
        ADC & 2.38 & 20.67 \\
        TCD & 2.40 & 20 \\
        TDC & 2.40 & 20 \\
        CTD & 2.30 & 23.33 \\
        CDT & 2.40 & 20 \\
        DTC & 2.28 & 24 \\
        DCT & 2.31 & 23 \\
        \bottomrule
    \end{tabular*}
\end{table}

\begin{figure}[htbp]
	\centering
	\includegraphics{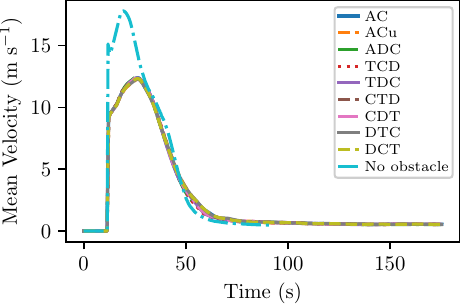}
	\caption{Effect of uniform and mixed-shape obstacle array configurations, along with the no-obstacle case, on the mean velocity of a snow avalanche near the roadside region at the CT-8 avalanche site. The simulation input parameters are: release area = 61,800 \si{\square\meter}, release height = 25 \si{\centi\meter}, internal friction angle $\phi = 15^\circ$, and bed friction angle $\delta = 27.5^\circ$.}
	\label{fig:ct_8_mean_velocity}
\end{figure}

Similarly, as depicted in Figure~\ref{fig:ct_8_mean_velocity}, the presence of the obstacle arrays reduced the (mass-weighted) mean avalanche velocity compared to the no-obstacle case. Specifically, the mean velocity of the flow reaching the roadside region decreased by approximately 25\%. Consistent with the flow depth observations, variations in obstacle shape had only a minimal effect on the mean velocity. 
These results suggest that, under the field-scale conditions examined, terrain characteristics have a greater influence on the predicted roadside flow depth and mean velocity than the variations in obstacle shape considered in this study.
Finally, the overall deceleration due to obstacle arrays is consistent with the findings of \citet{bi2019effects}, who demonstrated a similar reduction in flow velocity due to obstacle array at the Zhangmu avalanche site in Tibet.

\section{Conclusion}
\label{sec:7}
This study investigated the effects of obstacle arrays placed in the flow path of an avalanche using a finite-volume solver based on a depth-averaged approach. The numerical framework was first validated against small-scale experiments involving a configuration of three hemispherical obstacles. The validation demonstrated the model's ability to capture obstacle–flow interactions, including the time of jump, temporal evolution of flow depth, and steady-state flow depth.

First, the avalanche–obstacle interaction was investigated to examine the formation of dead zones and grain-free regions. Subsequently, the effects of obstacle height and width were analyzed in terms of blockage ratio, velocity reduction ratio, and steady-state depth extent. The results demonstrated that arrays with a uniform obstacle height of 0.06 \si{\meter} and a corresponding width of 0.048 \si{\meter} were the most effective in obstructing the flow. Next, the effects of obstacle geometry, specifically cylindrical, thin cuboid, and diamond cuboid shapes, were investigated using upstream flow deposition, blockage ratio, steady-state depth extent, and impact force as evaluation parameters. Comparisons between uniform and mixed-shape configurations revealed that arrays with diamond cuboid obstacles in the first row provided the most effective flow obstruction. In the small-scale setup, configurations with diamond cuboid obstacles in the leading row resulted in the highest upstream flow deposition and blockage ratio, the lowest steady-state depth extent, and the maximum impact force. Furthermore, empirical correlations were developed to relate the maximum flow depth, its spatial position, and the mean avalanche velocity corresponding to the position of maximum flow depth to obstacle height, width, and shape.

The investigation was further extended to a real-world CT-8 avalanche site in Jammu \& Kashmir, India, where both uniform and mixed-shape obstacle configurations were implemented. The numerical simulations demonstrated that the obstacle arrays effectively reduced flow depth and mean avalanche velocity in the roadside region. However, unlike the small-scale analysis, the influence of obstacle shape on the overall avalanche flow over complex mountain terrain was found to be marginal. Nevertheless, obstacle geometry and dimensions can play an important role in protecting structures located immediately downstream of the obstacles, where flow conditions and the extent of flow obstruction are critical to effective protection. In this context, the small-scale findings provide useful insights into selecting obstacle shapes and dimensions for configurations designed to meet specific protection requirements. Future work could incorporate different rheological models into the depth-averaged framework to further investigate obstacle–flow interactions under different flow regimes. Overall, the findings demonstrate the potential of obstacle arrays as a mitigation measure in avalanche-prone regions, particularly for protecting vulnerable infrastructure and reducing avalanche-related hazards.

\newpage
\setcounter{table}{0}
\renewcommand{\thetable}{A.\arabic{table}}
\appendix
\addcontentsline{toc}{section}{Appendix}
\section{Numerical data for empirical correlation}\label{app:empirical}
The following table represents the values of maximum flow depth, its spatial location, and the mean velocity value corresponding to the location of maximum flow depth, obtained from numerical simulations across all array configurations, for uniform and mixed obstacle height, width, and shapes. 

\begin{table}[width=\linewidth,cols=9,pos=h]
	\caption{Comparison of actual numerical results and predicted values for maximum flow depth ($h_{\max}$), spatial coordinates ($X_{\mathrm{cen}}$, $Y_{\mathrm{cen}}$), and mean velocity ($V_{\mathrm{mean}}$) across obstacle array configurations with varying heights, widths, and shapes.}
	\label{tab:numerical_data}
	\scriptsize
	\setlength{\tabcolsep}{2.5pt}
	\begin{tabular*}{\tblwidth}{@{\extracolsep{\fill}} l cccccccc @{}}
		\toprule
		\textbf{Config.} & \multicolumn{2}{c}{\textbf{$h_{\max}$} (\si{\meter})} & \multicolumn{2}{c}{\textbf{$X_{\mathrm{cen}}$} (\si{\meter})} & \multicolumn{2}{c}{\textbf{$Y_{\mathrm{cen}}$} (\si{\meter})} & \multicolumn{2}{c}{\textbf{$V_{\mathrm{mean}}$} (\si{\meter\per\second})} \\
		\cmidrule(lr){2-3} \cmidrule(lr){4-5} \cmidrule(lr){6-7} \cmidrule(lr){8-9}
		& Actual & Pred. & Actual & Pred. & Actual & Pred. & Actual & Pred. \\
		\midrule
		AC-OH1  & 0.053239 & 0.053599228 & 0.356482 & 0.3565256 & 0.499566 & 0.499554 & 1.02059  & 1.020311 \\
		AC-OH2  & 0.05171  & 0.054095011 & 0.357634 & 0.3575855 & 0.499564 & 0.499554 & 1.03197  & 1.031057 \\
		AC-OH3  & 0.058313 & 0.05448264  & 0.358273 & 0.3583361 & 0.499584 & 0.499554 & 1.04075  & 1.040312 \\
		AC-CS1  & 0.058313 & 0.05448264  & 0.358271 & 0.3583544 & 0.499584 & 0.499554 & 1.04033  & 1.039542 \\
		AC-CS2  & 0.05171  & 0.054095011 & 0.357634 & 0.3575156 & 0.499563 & 0.499554 & 1.03235  & 1.031814 \\
		AC-CS3  & 0.053239 & 0.053599228 & 0.356482 & 0.3565043 & 0.499565 & 0.499554 & 1.02107  & 1.021196 \\
		AC-CS4  & 0.058313 & 0.05448264  & 0.358271 & 0.3583536 & 0.499584 & 0.499554 & 1.03962  & 1.039576 \\
		AC-CS5  & 0.05171  & 0.054095011 & 0.357634 & 0.3575856 & 0.499563 & 0.499554 & 1.03134  & 1.031052 \\
		AC-CS6  & 0.053239 & 0.053599228 & 0.356478 & 0.356505  & 0.499566 & 0.499554 & 1.02136  & 1.021167 \\
		ACu-OH1 & 0.055331 & 0.056925442 & 0.317679 & 0.324968  & 0.49961  & 0.499599 & 1.03867  & 1.035162 \\
		ACu-OH2 & 0.058113 & 0.057451992 & 0.358144 & 0.3459711 & 0.499555 & 0.499569 & 1.02303  & 1.026543 \\
		ACu-OH3 & 0.060708 & 0.057863676 & 0.359298 & 0.366779  & 0.499559 & 0.499547 & 1.03277  & 1.027171 \\
		ACu-CS1 & 0.055331 & 0.057863676 & 0.317677 & 0.3269965 & 0.49961  & 0.499598 & 1.03872  & 1.033165 \\
		ACu-CS2 & 0.058113 & 0.057451992 & 0.358141 & 0.3464732 & 0.499553 & 0.499574 & 1.02233  & 1.034286 \\
		ACu-CS3 & 0.060708 & 0.056925442 & 0.359294 & 0.3717461 & 0.49956  & 0.499539 & 1.03208  & 1.02813  \\
		ACu-CS4 & 0.055331 & 0.057863676 & 0.317676 & 0.3286732 & 0.49961  & 0.499596 & 1.03876  & 1.032897 \\
		ACu-CS5 & 0.058113 & 0.057451992 & 0.358147 & 0.3457528 & 0.499554 & 0.499569 & 1.02364  & 1.026575 \\
		ACu-CS6 & 0.060708 & 0.056925442 & 0.35929  & 0.3701122 & 0.49956  & 0.499541 & 1.03124  & 1.02836  \\
		ADC-OH1 & 0.069701 & 0.068931139 & 0.491106 & 0.4907052 & 0.499314 & 0.499318 & 0.716326 & 0.713889 \\
		ADC-OH2 & 0.066847 & 0.068931139 & 0.496768 & 0.49614   & 0.499439 & 0.499347 & 0.752868 & 0.743822 \\
		ADC-OH3 & 0.069422 & 0.068532495 & 0.565057 & 0.5649018 & 0.499279 & 0.49927  & 0.561243 & 0.557726 \\
		ADC-CS1 & 0.069857 & 0.068532495 & 0.56431  & 0.5639438 & 0.499224 & 0.499251 & 0.555086 & 0.552667 \\
		ADC-CS2 & 0.067763 & 0.068532495 & 0.560681 & 0.5599839 & 0.499356 & 0.499246 & 0.546017 & 0.539906 \\
		ADC-CS3 & 0.07037  & 0.068931139 & 0.491265 & 0.4916817 & 0.499309 & 0.499341 & 0.719149 & 0.721549 \\
		ADC-CS4 & 0.068385 & 0.068532495 & 0.563038 & 0.5639865 & 0.499276 & 0.499251 & 0.5447   & 0.552892 \\
		ADC-CS5 & 0.067194 & 0.068931139 & 0.496439 & 0.4961354 & 0.499458 & 0.499347 & 0.747495 & 0.743784 \\
		ADC-CS6 & 0.069681 & 0.068931139 & 0.491368 & 0.4916497 & 0.499321 & 0.49934  & 0.720816 & 0.721296 \\
		CDT     & 0.053239 & 0.053599228 & 0.356484 & 0.3565256 & 0.499564 & 0.499554 & 1.02007  & 1.020311 \\
		CTD     & 0.053239 & 0.053599228 & 0.356479 & 0.3565256 & 0.499565 & 0.499554 & 1.01981  & 1.020311 \\
		DCT     & 0.06955  & 0.068931139 & 0.491295 & 0.4907052 & 0.499314 & 0.499318 & 0.720398 & 0.713889 \\
		DTC     & 0.069968 & 0.068931139 & 0.491061 & 0.4907052 & 0.499327 & 0.499318 & 0.716206 & 0.713889 \\
		TCD     & 0.055331 & 0.056925442 & 0.317674 & 0.324968  & 0.49961  & 0.499599 & 1.03871  & 1.035162 \\
		TDC     & 0.055331 & 0.056925442 & 0.317674 & 0.324968  & 0.49961  & 0.499599 & 1.03862  & 1.035162 \\
		\bottomrule
	\end{tabular*}
\end{table}

\section*{Funding}
The authors declare that no funds, grants, or other support were received during the preparation of this manuscript.

\printcredits

\section*{Declaration of competing interest}
The authors declare that they have no known competing financial interests or personal relationships that could have appeared to influence the work reported in this paper.

\section*{Acknowledgments}
VS would like to express his gratitude to the Ministry of Education (MoE), India, for providing him with a scholarship for his PhD studies.

\section*{Code availability}
The test case files for simulating avalanche-obstacle array interactions are publicly available at \url{https://github.com/Arshavin93/obstacle_test_case_files}.

\bibliographystyle{cas-model2-names}

\bibliography{cas-refs}

\end{document}